\documentclass[11pt]{article}

\usepackage[a4paper,margin=25mm]{geometry}
\usepackage{graphicx}
\usepackage{multirow}
\usepackage{amsmath,amssymb,amsfonts}
\usepackage{amsthm}
\usepackage{mathrsfs}
\usepackage[title]{appendix}
\usepackage{xcolor}
\usepackage{textcomp}
\usepackage{booktabs}
\usepackage{algorithm}
\usepackage{algorithmicx}
\usepackage{algpseudocode}
\usepackage{listings}
\usepackage{array}
\usepackage{tabularx}
\usepackage{enumitem}
\usepackage{subcaption}
\usepackage{placeins}
\usepackage[authoryear,round]{natbib}
\usepackage[hidelinks]{hyperref}
\usepackage{authblk}

\theoremstyle{plain}

\theoremstyle{definition}

\theoremstyle{remark}

\title{Goal-Oriented Weighting of Reynolds-Stress Data for Learning Turbulence Models in Complex Flows}

\author[1,2]{Zhuolin Zhao}
\author[1,2]{Haochen Wang}
\author[3]{Youngwoo Kim}
\author[3]{Solkeun Jee\thanks{Corresponding author: \texttt{sjee@gist.ac.kr}}}
\author[1,2]{Heng Xiao\thanks{Corresponding author: \texttt{heng.xiao@simtech.uni-stuttgart.de}}}

\affil[1]{Stuttgart Center for Simulation Science, University of Stuttgart, Stuttgart 70569, Germany}
\affil[2]{Institute of Aerospace Thermodynamics, University of Stuttgart, Stuttgart 70569, Germany}
\affil[3]{Department of Mechanical and Robotics Engineering, Gwangju Institute of Science and Technology, Gwangju 61005, South Korea}
\date{}

\begin{document}
\maketitle

\begin{abstract}
Data-driven turbulence models for the Reynolds-averaged Navier--Stokes (RANS) equations offer a promising route to improving predictions of complex flows. Such models, specifically the turbulence constitutive relations, can be learned efficiently from high-fidelity Reynolds-stress data without evaluating the RANS equations during training. However, conventional losses weight all tensor-component and spatial errors equally, although their influence on a quantity of interest (QoI) can differ significantly; reducing the aggregate stress error therefore does not necessarily lead to an improved QoI prediction. Training against flow-level observations accounts for this dependence but requires repeated, potentially expensive RANS solutions. In canonical shear flows, physical reasoning can identify the shear component as the only one relevant for the mean-flow prediction. Motivated by this example, we propose a goal-oriented method to select and weight Reynolds-stress training data for complex flows. For a specified QoI, a single offline adjoint evaluation quantifies its sensitivity to local perturbations of each Reynolds-stress component. These sensitivities are converted into fixed weights in the supervised loss, enabling training without further RANS solutions. We validate the weighting in square-duct and periodic-hill flows, then apply it to a film-cooling jet in crossflow, where it improves velocity and cooling-effectiveness predictions over uniformly weighted training despite using a velocity-only QoI. Beyond turbulence modelling, this work suggests a broader strategy for aligning supervised learning with downstream prediction goals while preserving training efficiency.
\end{abstract}


\section{Introduction}\label{intro}

The Reynolds-averaged Navier--Stokes (RANS) equations remain the workhorse of engineering flow simulation, but their predictive accuracy depends on modelling the effects of unresolved turbulent fluctuations. These effects enter the mean momentum equations through the Reynolds stress \(\boldsymbol{\tau}\), making its deviatoric part a central target of turbulence closure~\citep{pope2000turbulent}. Conventional linear eddy-viscosity models assume that the deviatoric part of the Reynolds stress, \(\operatorname{dev}(\boldsymbol{\tau})\), is aligned with the mean strain-rate tensor \(\mathbf{S}\) and related to it through a scalar eddy viscosity \(\nu_t\), i.e., \(\operatorname{dev}(\boldsymbol{\tau}) = 2\nu_t\mathbf{S}\). This assumption cannot represent stress anisotropy that is not proportional to the local mean strain, including important normal-stress imbalances and stress--strain misalignment, thereby limiting predictive accuracy in flows involving separation, streamline curvature, and stress-driven secondary motion \citep{pope2000turbulent,xiao2019quantification}. Data-driven closures seek to overcome this representational limitation by learning more expressive constitutive relations from high-fidelity flow data. High-fidelity information can enter this learning process at two different levels. \emph{Direct training} fits the constitutive relation \(\operatorname{dev}(\boldsymbol{\tau}) = \mathbf{f}(\mathbf{S}, \boldsymbol{\Omega})\), with \(\boldsymbol{\Omega}\) the mean rotation-rate tensor, against closure-level data, i.e., the Reynolds stresses~\citep{ling2016reynolds,wang2017physics,wu2018physics,weatheritt2016novel,schmelzer2020discovery,cherroud2025space}. In contrast, \emph{indirect training} infers the closure from flow-level observations, such as velocity or aerodynamic forces, by propagating its effect through the RANS equations \citep{duraisamy2019turbulence,parish2016paradigm,volpiani2021machine,michelenstrofer2021endtoend,zhang2022ensemble,zhao2020rans,saidi2022cfd}. The direct and indirect training strategies are also described, respectively, as \emph{a priori} training~\citep{duraisamy2021perspectives} and as model-consistent learning~\citep{duraisamy2021perspectives,zhang2022ensemble} or CFD-driven learning~\citep{zhao2020rans}.

Direct training provides a computationally efficient route for learning constitutive models from high-fidelity Reynolds-stress data~\citep{duraisamy2021perspectives}.
The Reynolds stresses represent the unclosed turbulent momentum transport in the RANS equations and therefore provide a natural target for supervised closure learning. In the standard two-stage procedure, fields from direct numerical simulation (DNS) or large-eddy simulation (LES) are first processed to obtain closure-level targets, such as Reynolds-stress anisotropy, stress discrepancies, or constitutive coefficients, and a model is then trained offline to predict the selected target from local flow features \citep{ling2016reynolds,wang2017physics,weatheritt2016novel,schmelzer2020discovery,kaandorp2020data}. After training, the model parameters remain fixed, and the constitutive relation is evaluated using the evolving RANS flow state. Because the governing equations are not solved during parameter optimisation, this procedure is much cheaper than solver-in-the-loop training. A common supervised loss aggregates pointwise reconstruction errors with uniform nominal weights over anisotropy components and sampled locations, often at each cell centre~\citep{ling2016reynolds,schmelzer2020discovery}. Closure errors are subsequently propagated through, and can be amplified by, the RANS equations \citep{wu2019reynolds,brener2021conditioning}. Because individual component--location errors can have different effects on a selected flow quantity of interest (QoI), such as the velocity at sampling points, reducing their aggregate reconstruction error does not necessarily improve the intended flow prediction. Moreover, the closure is trained in an environment different from the one in which it is deployed, and this inconsistency between the training and prediction environments can further degrade the \emph{a posteriori} predictions~\citep{duraisamy2021perspectives,zhang2022ensemble}.

Indirect training can account for this unequal influence by evaluating closure changes through the RANS equations and comparing the prediction with flow-level observations.
It is commonly implemented in a solver-in-the-loop form, in which velocity, aerodynamic forces, or another selected flow quantity defines the optimisation objective. The governing equations then provide the connection between changes in the closure and changes in the observable response. Different implementations obtain this connection in different ways. Adjoint-based field inversion first infers a spatial correction from flow observations and can subsequently learn a constitutive mapping from that correction \citep{parish2016paradigm}; end-to-end differentiable methods instead optimise the constitutive-model parameters directly through the solver \citep{michelenstrofer2021endtoend}. Ensemble-based methods estimate parameter--observation relationships from repeated forward simulations \citep{zhang2022ensemble,liu2026toward,luo2026physics}, while CFD-driven model-discovery methods rank candidate closures using their converged RANS predictions \citep{zhao2020rans}. These approaches can align model development with the intended flow prediction, but they require repeated flow solutions, solver sensitivities, or ensembles of simulations during optimisation and can therefore be computationally expensive. This computational burden motivates us to pursue a direct-training strategy that uses QoI relevance to prioritise Reynolds-stress data without repeated RANS solutions during model optimisation.

\begin{figure}[htbp]
    \centering
    \includegraphics[width=0.6\textwidth]{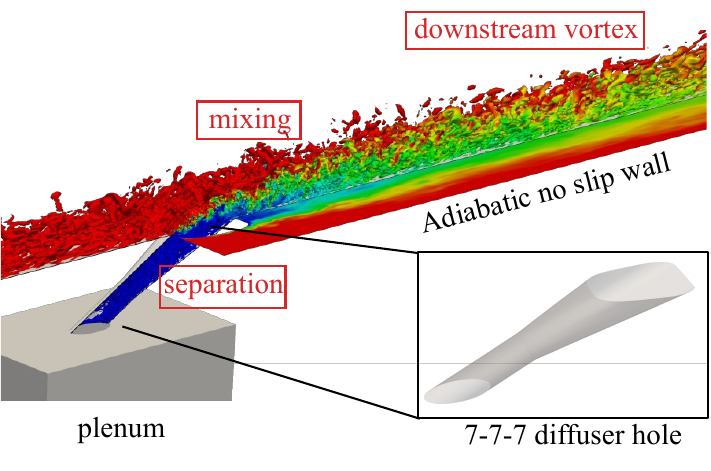}
    \caption{Flow mechanisms associated with spatially varying Reynolds-stress relevance in the \(7\text{-}7\text{-}7\) jet-in-crossflow configuration. Coolant enters from the plenum, passes through the fan-shaped diffusing hole, and interacts with the mainstream near the hole exit}
    \label{fig:jic-motivation}
\end{figure}

The need to prioritise Reynolds-stress data according to the intended prediction can be understood from canonical shear flows. In fully developed plane-channel flow, the Reynolds shear stress is the only Reynolds-stress component entering the streamwise mean-momentum balance \citep{pope2000turbulent}. A model trained for such a flow should therefore fit the shear stress accurately, whereas errors in the normal stresses do not affect the mean velocity. This example suggests a more general principle: Reynolds-stress components and spatial locations should receive attention according to their influence on the selected flow quantity. These priorities become more difficult to identify in complex three-dimensional flows, where several coupled mechanisms impose different modelling requirements across the domain. Such a challenge is most clearly exemplified by the jet in crossflow, in which a jet is injected into a crossflow, e.g., to cool turbine blades \citep{bunker2005review}. In the film-cooling configuration of Fig.~\ref{fig:jic-motivation}, coolant from the plenum passes through a fan-shaped diffusing hole, where the flow may separate \citep{Schroeder2014Baseline,Gunady2021Velocity}, and exchanges momentum with the crossflow near the hole; further downstream, vortical structures, most notably the counter-rotating vortex pair, mix the coolant with the mainstream \citep{mahesh2013interaction}. These mechanisms occupy distinct regions and can involve different Reynolds-stress components. Because these mechanisms redistribute momentum in different directions and regions, the stress components most relevant to a selected velocity QoI can vary spatially. Data-driven Reynolds-stress anisotropy modelling has also been investigated for film and effusion cooling \citep{ellis2023data}. For such flows, uniform weighting is not merely inefficient; it defines a training loss that is structurally misaligned with the intended prediction. A systematic scheme for identifying the Reynolds-stress data most influential to that prediction is therefore required. Related work accounts for spatially varying model performance by learning weights that combine predictions from competing turbulence models \citep{de2024space,cherroud2025space}. In the present work, the weights instead act on Reynolds-stress component errors in the training loss and are determined from their sensitivity to the selected QoI.

We propose a goal-oriented weighting framework that embeds the influence of Reynolds-stress data on a selected flow quantity into direct turbulence-model training (Section~\ref{sec2}). For each training flow, a single offline adjoint solution of the baseline RANS equations quantifies the local linearised sensitivity of a velocity-based QoI to perturbations in each Reynolds-stress anisotropy component. These sensitivities are converted into fixed weights over tensor components and spatial locations in the supervised loss, directing the model toward the stress data with greater influence on the intended prediction. When the relative component priorities are approximately uniform across the domain, they can be represented by global component weights; when they vary spatially, cellwise component weights retain the local sensitivity information. The flow and adjoint solvers are used only to construct the weighted training loss; once the fixed weights have been obtained, model optimisation proceeds using the same supervised learning procedure as conventional direct training, without further RANS evaluations.

The weighting principle is first validated on two canonical flows, the flow in a square duct and the flow over periodic hills, in which the stress mechanisms governing secondary motion and flow separation are well established. It is then applied to the jet in crossflow, a complex three-dimensional configuration in which the relevant stress components vary in space.

\FloatBarrier
\section{Methodology}\label{sec2}

\begin{figure}[htbp]
    \centering
    \includegraphics[width=0.9\textwidth]{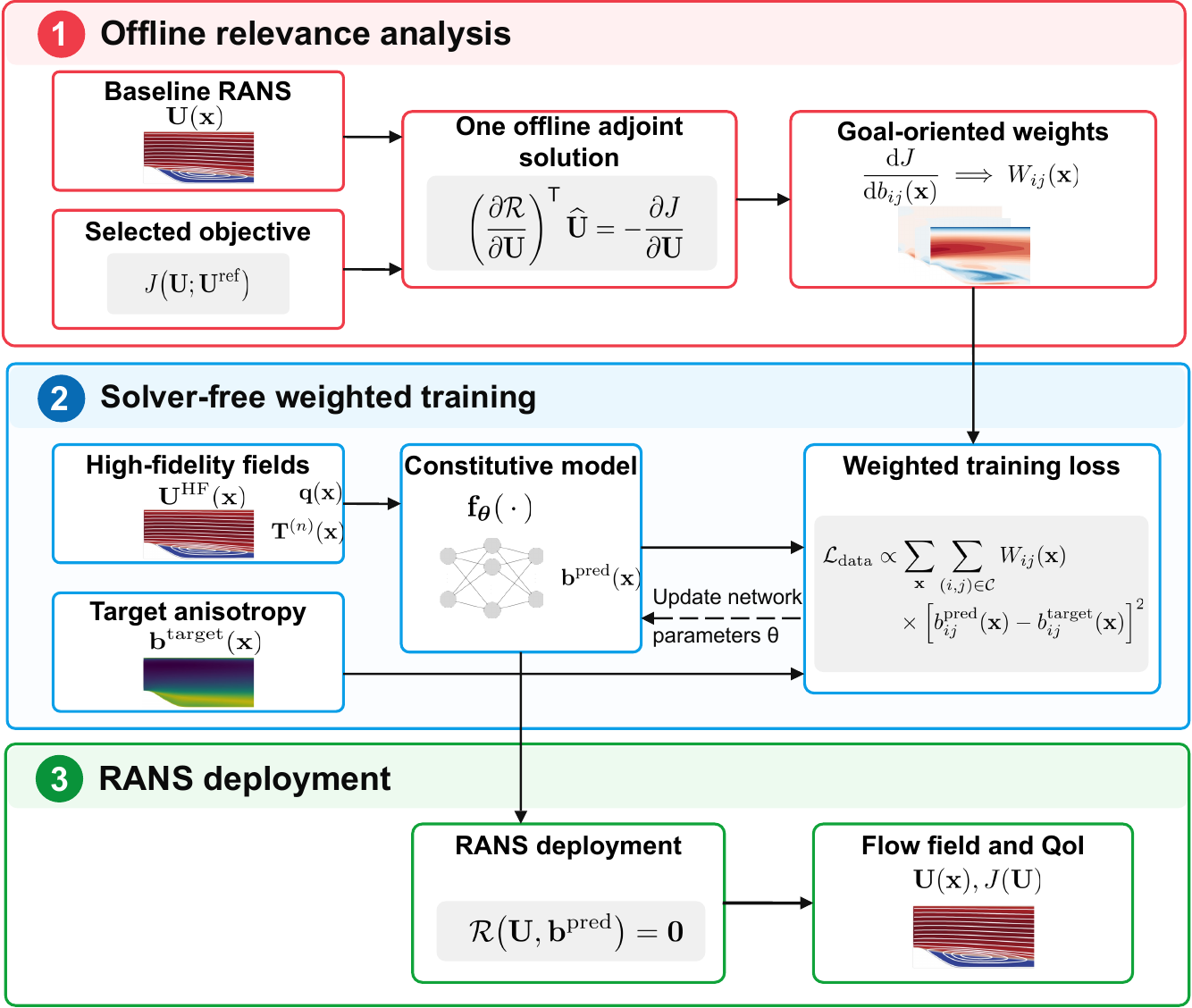}
    \caption{Workflow of goal-oriented weighting for direct turbulence-model training. The selected velocity QoI and baseline RANS state define a single offline adjoint calculation, whose anisotropy sensitivities are converted into fixed spatial and component weights. These weights prioritise the high-fidelity constitutive data during solver-free supervised training. The resulting constitutive model is subsequently deployed in the RANS solver to obtain the flow and QoI predictions. Here \(\mathcal{R}\) denotes the RANS residual operator and \(\boldsymbol{\theta}\) the parameters of the constitutive model; the remaining symbols are introduced in Sections~\ref{sec2} and \ref{sec:weighting} as they arise}
    \label{fig:method-framework}
\end{figure}

The proposed framework introduces quantity-of-interest relevance into direct turbulence-model training while preserving solver-free parameter optimisation. As summarised in Fig.~\ref{fig:method-framework}, a selected velocity QoI and the corresponding baseline RANS state are used in a single offline adjoint calculation to identify the influential anisotropy components and spatial locations. The resulting sensitivities are converted into fixed loss weights that prioritise the high-fidelity constitutive data during supervised training, after which the learned constitutive relation is deployed in the RANS solver to obtain the flow and QoI predictions.

\subsection{Direct data-training framework}
\label{sec:direct-training}

The direct-training framework learns a constitutive model for the Reynolds-stress anisotropy from high-fidelity data and subsequently deploys the learned closure in the RANS equations. For a statistically stationary incompressible flow, the governing equations are
\begin{equation}
\nabla\!\cdot\mathbf{U}=0,
\qquad
(\mathbf{U}\!\cdot\!\nabla)\mathbf{U}
=
-\nabla p
+\nu\nabla^2\mathbf{U}
+\nabla\!\cdot\boldsymbol{\tau},
\label{eq:rans}
\end{equation}
where \(\mathbf{U}\), \(p\), and \(\nu\) denote the mean velocity, kinematic pressure, and molecular kinematic viscosity, respectively, and \(\boldsymbol{\tau}=-\langle\mathbf{u}'\mathbf{u}'\rangle\) is the Reynolds-stress tensor, with \(\mathbf{u}'\) the velocity fluctuation and \(\langle\cdot\rangle\) the Reynolds average. With the turbulent kinetic energy defined as \(k=-\tfrac{1}{2}\operatorname{tr}(\boldsymbol{\tau})\), the Reynolds stress is decomposed as
\begin{equation}
\boldsymbol{\tau}
=
-\frac{2}{3}k\mathbf{I}
-
2k\mathbf{b},
\label{eq:anisotropy}
\end{equation}
where \(\mathbf{b}\) is the symmetric, dimensionless anisotropy tensor \citep{pope2000turbulent}. Equations~(\ref{eq:rans}) and (\ref{eq:anisotropy}) are presented in incompressible form only to introduce this closure; the jet-in-crossflow calculations employ the corresponding compressible RANS formulation together with the energy equation.

The anisotropy is represented using an invariant tensor basis,
\begin{equation}
\mathbf{b}^{\mathrm{pred}}
=
\sum_{n=1}^{4}
g^{(n)}\mathbf{T}^{(n)},
\label{eq:present-tbnn}
\end{equation}
where \(\mathbf{T}^{(n)}\) are the first four tensors retained from the general ten-term integrity basis and \(g^{(n)}\) are scalar coefficient functions \citep{pope1975more}. In the present implementation, a tensor-basis neural network (TBNN) maps an invariant feature vector \(\mathbf{q}\) to these coefficients, \(g^{(n)}=g^{(n)}(\mathbf{q})\) \citep{ling2016reynolds}. This construction is invariant to a uniform change in inertial reference velocity and transforms the predicted anisotropy consistently under coordinate rotation. The feature selection and normalisation follow our previous work \citep{liu2026toward}; the complete feature definitions, tensor bases, case-specific input selection, and normalisation procedure are provided in Appendix~\ref{app:tbnn-details}. The proposed weighting modifies the training loss; the closure representation remains unchanged.

Direct training often calibrates the constitutive relation within a consistent high-fidelity flow state. The input features and tensor bases are evaluated using high-fidelity mean-flow quantities, and the target used in the square-duct and periodic-hill cases is the physical high-fidelity anisotropy,
\begin{equation}
\mathbf{b}^{\mathrm{HF}}
=
-\frac{\operatorname{dev}\!\left(\boldsymbol{\tau}^{\mathrm{HF}}\right)}
{2k^{\mathrm{HF}}},
\label{eq:hf-anisotropy-target}
\end{equation}
where \(\operatorname{dev}(\mathbf{A})=\mathbf{A}-\tfrac{1}{3}\operatorname{tr}(\mathbf{A})\mathbf{I}\) denotes the deviatoric part of a tensor \(\mathbf{A}\). Because the model is subsequently deployed with RANS mean-flow inputs, this procedure retains the inconsistency between training and prediction environments noted in Section~\ref{intro}; the weighting proposed in this work addresses the unequal influence of the data on the QoI, not this inconsistency.

For the jet-in-crossflow case, an effective anisotropy target is used to account for the mismatch between the LES and RANS turbulent kinetic energies when reconstructing the deviatoric Reynolds stress. \citet{ellis2023data} identified this mismatch as a limitation of predictions obtained with learned anisotropy and examined \emph{a posteriori} calculations with a prescribed LES kinetic-energy field. The present work follows their diagnosis but not their remedy: instead of prescribing the high-fidelity kinetic energy at deployment, the stress-scale discrepancy is incorporated into the training target:
\begin{equation}
\mathbf{b}^{\mathrm{eff}} = -\frac{\operatorname{dev}\!\left(\boldsymbol{\tau}^{\mathrm{LES}}\right)}{2k^{\mathrm{RANS}}},
\label{eq:jic-effective-anisotropy}
\end{equation}
where \(k^{\mathrm{RANS}}\) in this definition is the baseline-RANS kinetic-energy field at the training condition. During \emph{a posteriori} deployment, the turbulent kinetic energy continues to evolve through the RANS turbulence-transport equations rather than being prescribed from LES, allowing the learned closure to be applied under other operating conditions for which high-fidelity \(k\) data are unavailable.

With these constitutive targets defined, conventional direct training fits all available component--cell data with equal nominal importance. The following subsection replaces this uniform treatment with weights that reflect the influence of each anisotropy component and spatial location on the selected QoI.

\subsection{Adjoint-based QoI-sensitivity weighting}

\label{sec:weighting}

The proposed weighting modifies the supervised training loss so that anisotropy errors receive different emphasis according to their influence on the selected flow prediction. Each component--cell error is assigned a weight \(W_{ij}(\mathbf{x})\), constructed from one offline adjoint sensitivity analysis for each training flow and selected objective. These weights remain fixed during model optimisation: the flow-level objective guides their construction, while the weighted anisotropy-fitting loss is minimised during training.

The resulting training loss is
\begin{equation}
\mathcal{L}_{\mathrm{data}}=
\frac{1}{N_{\mathcal{F}}N_c}
\sum_{\mathbf{x}\in\Omega_{\mathrm{map}}}
\sum_{(i,j)\in\mathcal{C}}
W_{ij}(\mathbf{x})
\left[b_{ij}^{\mathrm{pred}}(\mathbf{x})-b_{ij}^{\mathrm{target}}(\mathbf{x})\right]^2,
\label{eq:weighted-training-loss}
\end{equation}
where \(\Omega_{\mathrm{map}}\) contains the RANS cells onto which the high-fidelity data are mapped, \(\mathcal{C}\) contains the \(N_c=6\) unique symmetric anisotropy components, and \(N_{\mathcal{F}}\) is the number of cells in the retained training region \(\mathcal{F}\). The weights vanish outside the retained region \(\mathcal{F}\); within it, a larger weight assigns greater emphasis to the corresponding fitting error. The target anisotropies are defined in Eqs.~\eqref{eq:hf-anisotropy-target} and \eqref{eq:jic-effective-anisotropy}. Retaining all mapped cells and setting \(W_{ij}=1\) recovers uniformly weighted training.

The weight separates spatial selection, cell relevance, and component priority into three factors:
\begin{equation}
W_{ij}(\mathbf{x})=M(\mathbf{x})\,c(\mathbf{x})\,w_{ij}(\mathbf{x}),
\label{eq:cellwise-adjoint-weight}
\end{equation}
where the binary mask \(M(\mathbf{x})\) determines whether a cell participates in training, the cell-relevance factor \(c(\mathbf{x})\) controls the relative emphasis among retained cells, and the component factor \(w_{ij}(\mathbf{x})\) distributes that emphasis among the anisotropy components within each cell. The factors \(c\) and \(w_{ij}\) are extended by zero outside \(\mathcal{F}\). Their construction uses the sensitivity of the selected velocity objective to local anisotropy perturbations.

To construct these factors, we first define the velocity objective used to assess data relevance. For a selected velocity component \(U_m\) at \(N_s\) sampling locations \(\mathbf{x}_n\), the objective is
\begin{equation}
J=\frac{1}{2N_s}\sum_{n=1}^{N_s}\left[U_m(\mathbf{x}_n)-U_m^{\mathrm{ref}}(\mathbf{x}_n)\right]^2,
\label{eq:qoi}
\end{equation}
where \(U_m^{\mathrm{ref}}\) denotes the corresponding high-fidelity reference. The sensitivity of \(J\) is evaluated about the converged baseline RANS solution at the training condition. One offline adjoint solution then determines how this objective responds to a local perturbation of each anisotropy component \citep{giles2000introduction}. We define the corresponding component sensitivity magnitude as
\begin{equation}
G_{ij}(\mathbf{x})=
\left|
\frac{\mathrm{d} J}{\mathrm{d} b_{ij}(\mathbf{x})}
\right|,
\qquad
(i,j)\in\mathcal{C},
\label{eq:absolute-sensitivity}
\end{equation}

where each off-diagonal entry represents the paired symmetric components \(b_{ij}=b_{ji}\). This construction can be considered a frozen counterpart of end-to-end differentiable training. In that approach the model parameters are updated with \(\mathrm{d}J/\mathrm{d}w=(\mathrm{d}J/\mathrm{d}\boldsymbol{\tau})(\mathrm{d}\boldsymbol{\tau}/\mathrm{d}w)\), where the sensitivity \(\mathrm{d}J/\mathrm{d}\boldsymbol{\tau}\) is obtained from an adjoint solution recomputed at every parameter update \mbox{\citep{michelenstrofer2021endtoend}}. The present scheme evaluates the same sensitivity only once at the baseline RANS solution, and uses it as fixed weights without involving the adjoint solver during training. Larger values of \(G_{ij}(\mathbf{x})\) identify the components and locations with greater local influence on the selected velocity objective. These sensitivities provide the input for constructing the spatial mask \(M(\mathbf{x})\), the cell-relevance factor \(c(\mathbf{x})\), and the component-priority factor \(w_{ij}(\mathbf{x})\). The adjoint equations, the projection onto the symmetric and traceless anisotropy representation, and the treatment of paired off-diagonal components are detailed in Appendix~\ref{app:weights}.

\begin{figure}[htbp]
    \centering
    \includegraphics[width=0.5\linewidth]{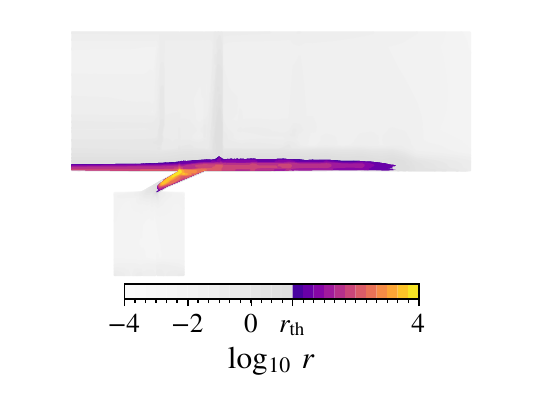}
    \caption{Adjoint-based selection of the fit region for the jet-in-crossflow case. The normalised cell sensitivity \(r\) is shown together with the cells retained for direct training}
    \label{fig:jic-mask-selection}
\end{figure}

\textit{Spatial selection.} The mask focuses training on cells with appreciable sensitivity to the selected objective. In the jet-in-crossflow case, many cells in the freestream and plenum have little influence on the downstream velocity objective, yet fitting their stress targets can compete with fitting the more influential injection and jet-interaction regions. To identify the relevant cells, the component sensitivity magnitudes are averaged within each cell and normalised by the median over the mapped domain:
\begin{equation}
s(\mathbf{x})=\frac{1}{N_c}\sum_{(i,j)\in\mathcal{C}}G_{ij}(\mathbf{x}),
\qquad
r(\mathbf{x})=\frac{s(\mathbf{x})}{\displaystyle\operatorname{median}_{\mathbf{x}\in\Omega_{\mathrm{map}}}s(\mathbf{x})}.
\label{eq:cell-sensitivity-score}
\end{equation}
The normalised score \(r\) expresses cell relevance relative to the domain median, removing dependence on an overall positive scaling of the objective. Cells satisfying \(r(\mathbf{x})\ge r_{\mathrm{th}}\) form the retained region \(\mathcal{F}\), with \(M(\mathbf{x})=1\) inside this region and zero elsewhere. For the jet-in-crossflow case, the threshold is selected from the sensitivity distribution, retaining cells around the injection passage and downstream coolant--crossflow interaction (Fig.~\ref{fig:jic-mask-selection}). Appendix~\ref{app:weights} provides the threshold-selection criterion, its value, and the supporting histogram.

\textit{Spatial relevance.} Among the retained cells, greater aggregate sensitivity receives greater fitting emphasis through the cell-relevance factor:
\begin{equation}
c(\mathbf{x})=\frac{\sqrt{s(\mathbf{x})}}{\left\langle\sqrt{s}\right\rangle_{\mathcal{F}}},
\qquad \mathbf{x}\in\mathcal{F},
\label{eq:cell-magnitude-factor}
\end{equation}
where \(\langle\cdot\rangle_{\mathcal{F}}\) denotes the arithmetic mean over retained cells. The square root compresses the sensitivity range while preserving its ordering, moderating the contrast between highly sensitive cells and the rest of the retained region. Normalisation gives \(c\) a unit mean, so it redistributes spatial emphasis without introducing an arbitrary overall weight scale.

\textit{Component priority.} Within each retained cell, the component weights distribute this emphasis according to the relative sensitivities of the anisotropy components. A preliminary weighting factor \(\omega_{ij}\), unrelated to the specific dissipation rate \(\omega\), is constructed from the sensitivity relative to the largest component value and then normalised:
\begin{equation}
\begin{aligned}
\omega_{ij}(\mathbf{x})
&=\varepsilon+(1-\varepsilon)
\left[
\frac{G_{ij}(\mathbf{x})}
{\displaystyle\max_{(k,l)\in\mathcal{C}}G_{kl}(\mathbf{x})}
\right]^{\gamma_w},\\
w_{ij}(\mathbf{x})
&=\frac{N_c\,\omega_{ij}(\mathbf{x})}
{\displaystyle\sum_{(k,l)\in\mathcal{C}}\omega_{kl}(\mathbf{x})},
\qquad \mathbf{x}\in\mathcal{F}.
\end{aligned}
\label{eq:cellwise-component-priority}
\end{equation}
The floor \(\varepsilon=0.05\) preserves a nonzero fitting penalty for weakly sensitive components, while \(\gamma_w=1\) gives a linear dependence of the preliminary factor on relative sensitivity. Normalisation gives \(w_{ij}\) a unit mean across components, leaving the overall cell emphasis controlled by \(c\). Together, these normalisations give \(W_{ij}\) a unit mean over retained cells and components. The local construction allows component priorities to vary between flow regions.

For the square-duct and periodic-hill validation cases, the component priorities are approximately consistent across the domain and are represented by global component weights. Their construction is detailed in Appendix~\ref{app:weights}, and the resulting values are reported in Table~\ref{tab:global-adjoint-weights}.

\FloatBarrier
\subsection{Flow configurations and evaluation}
\label{sec:flow-configurations}

The framework is assessed using two controlled validation cases and one complex flow. The square-duct and periodic-hill flows test the weighting principle for stress-driven secondary motion and flow separation, respectively. The \(7\text{-}7\text{-}7\) diffuser jet in crossflow serves as the principal application, in which the QoI relevance of the anisotropy data varies across both spatial locations and tensor components. For each configuration, the uniformly weighted and adjoint-weighted models use identical source dataset, model architecture, and optimisation settings; only the training weights differ.

\begin{figure}[htbp]
    \centering

\begin{minipage}[c]{0.48\textwidth}
        \centering

        \begin{subfigure}[t]{\linewidth}
            \centering
            \includegraphics[width=0.85\linewidth]{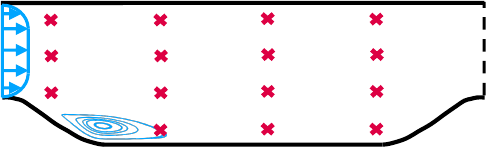}
            \caption{Periodic hill: \(U_x\) sampling points}
            \label{fig:flow-config-pehill}
        \end{subfigure}

        \vspace{0.8em}

        \begin{subfigure}[t]{\linewidth}
            \centering
            \includegraphics[width=0.85\linewidth]{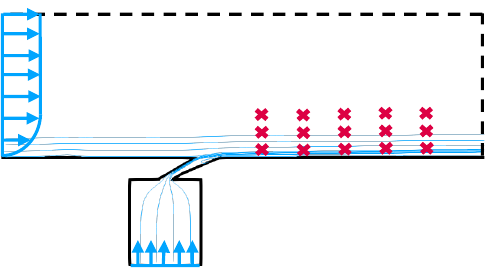}
            \caption{Jet-in-crossflow: \(U_x\) sampling points}
            \label{fig:flow-config-jic}
        \end{subfigure}
    \end{minipage}
    \hfill
\begin{subfigure}[c]{0.48\textwidth}
        \centering
        \includegraphics[width=0.95\linewidth]{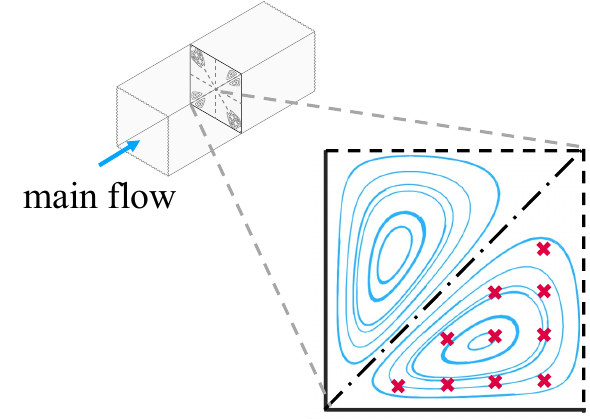}
        \caption{Square duct: \(U_y\) sampling points}
        \label{fig:flow-config-duct}
    \end{subfigure}

    \caption{Flow configurations and sampling points used to define the velocity-based QoIs. \textbf{\textcolor{red}{\texttimes}} denotes sparse sampling points for
    (a) the streamwise velocity in the periodic-hill flow, (b) the
    streamwise velocity in the jet-in-crossflow configuration, and (c) the in-plane velocity in the square duct. Blue arrows indicate the prescribed inflows, and the schematic contours identify the principal flow structures associated with each objective}
    \label{fig:flow-configurations}
\end{figure}

The velocity objective for each configuration is selected to represent the flow response examined in the deployed prediction. The in-plane velocity \(U_y\) is used for the square duct to target stress-driven secondary motion, while the streamwise velocity \(U_x\) is used for the periodic hill to target the separated shear layer and reversed-flow region. For the jet in crossflow, \(U_x\) targets the streamwise momentum redistribution associated with separation inside the diffusing hole and the subsequent development of the injected jet. Thermal quantities are not included in the adjoint objective; film-cooling effectiveness \(\eta\) is instead evaluated \emph{a posteriori} as a downstream response of the coupled momentum--scalar system. The normalised temperature is defined as \(\eta=(T_\infty-T)/(T_\infty-T_c)\), where \(T_\infty\) and \(T_c\) are the mainstream and coolant reference temperatures, respectively. At the adiabatic wall, \(T\) is the wall temperature, and \(\eta\) gives the adiabatic film-cooling effectiveness. The sampling points used to define the three velocity objectives are shown in Fig.~\ref{fig:flow-configurations}.

The training and test conditions are selected to assess generalisation within the two validation families and across operating conditions in the jet-in-crossflow application, as summarised in Table~\ref{tab:training-evaluation}. The square-duct model is trained using DNS data at \(\mathrm{Re}_{\mathrm{b}}=2600\) and tested at \(\mathrm{Re}_{\mathrm{b}}=1100\) and \(1800\) \citep{pinelli2010reynolds}, where \(\mathrm{Re}_{\mathrm{b}}=U_bh/\nu\), \(U_b\) is the bulk velocity, and \(h\) is the duct half-width. The periodic-hill model is trained using DNS data at the geometry parameter \(\alpha=1.5\) and tested at \(\alpha=1.0\) and \(1.2\) \citep{xiao2020flows}. The jet-in-crossflow model is trained using LES data for the \(7\text{-}7\text{-}7\) diffuser geometry at \(\mathrm{BR}=1.5\) and applied to the same geometry at \(\mathrm{BR}=1.0\) \citep{Schroeder2014Baseline}. The blowing ratio is defined as \(\mathrm{BR}=\rho_cU_c/(\rho_\infty U_\infty)\), where \(\rho_c\) and \(U_c\) are the coolant density and bulk velocity, respectively, and \(\rho_\infty\) and \(U_\infty\) are the corresponding mainstream quantities. The diameter \(D\) of the cylindrical inlet section of the cooling hole is used to normalise spatial coordinates. At \(\mathrm{BR}=1.0\), cooling effectiveness is compared with the experimental reference. The test conditions are excluded from both model training and weight construction. The square-duct and periodic-hill baselines use the \(k\)--\(\omega\) model,
whereas the jet-in-crossflow baseline uses the SST-2003 variant of the
\(k\)--\(\omega\) shear-stress-transport model.

\begin{table}[htbp]
    \centering
    \small
    \renewcommand{\arraystretch}{1.1}
    \setlength{\tabcolsep}{4pt}
    \begin{tabularx}{\textwidth}{@{}
        >{\raggedright\arraybackslash}p{0.18\textwidth}
        >{\raggedright\arraybackslash}X
        >{\centering\arraybackslash}p{0.18\textwidth}
        >{\centering\arraybackslash}p{0.20\textwidth}
        >{\centering\arraybackslash}p{0.12\textwidth}
        @{}}
        \toprule
        Configuration & Evaluation purpose & Training condition & Test condition & QoI \\
        \midrule
        Square duct
        & Controlled validation
        & \(\mathrm{Re}_{\mathrm{b}}=2600\)
        & \(\mathrm{Re}_{\mathrm{b}}=1100,\,1800\)
        & \(U_y\) \\

        Periodic hill
        & Controlled validation
        & \(\alpha=1.5\)
        & \(\alpha=1.0,\,1.2\)
        & \(U_x\) \\

        Jet in crossflow \(7\text{-}7\text{-}7\)
        & Principal application
        & \(\mathrm{BR}=1.5\)
        & \(\mathrm{BR}=1.0\)
        & \(U_x\) \\
        \bottomrule
    \end{tabularx}
    \caption{Training and test conditions for the two validation flows and the jet-in-crossflow application. Adjoint sensitivities and loss weights are constructed using only the corresponding training condition. Here, \(\mathrm{Re}_{\mathrm{b}}\) is the Reynolds number based on the bulk velocity, \(\alpha\) the slope parameter of the hill geometry, \(\mathrm{BR}\) the blowing ratio, \(U_x\) the streamwise velocity, and \(U_y\) the in-plane velocity.}
    \label{tab:training-evaluation}
\end{table}

The evaluation connects the learned anisotropy with the flow response obtained after deployment. For the square-duct and periodic-hill validation cases, \emph{a priori} comparisons focus on the anisotropy components identified as influential by the adjoint. For the jet in crossflow, the spatially varying component predictions are examined within the retained fit region; no single domain-averaged score is used. \emph{A posteriori} performance is assessed by deploying each trained closure in the RANS solver and comparing the resulting predictions with the available high-fidelity or experimental references. The evaluated quantities comprise secondary motion in the square duct, separation and reversed flow over the periodic hill, and velocity development and cooling effectiveness in the jet in crossflow.

\FloatBarrier
\section{Results}
Prioritising QoI-relevant Reynolds-stress data improves the deployed flow predictions compared with uniformly weighted training in the cases examined. The square-duct and periodic-hill cases validate the weighting principle against established stress mechanisms (Section~\ref{sec:controlled-validation}), while the jet in crossflow provides the principal assessment of spatially varying priorities (Section~\ref{sec:results-jic}). Generalisation is evaluated separately through changes in duct Reynolds number, hill geometry, and jet blowing ratio, without retraining the closures or recomputing the training weights.

\subsection{Controlled validation of goal-oriented weighting}
\label{sec:controlled-validation}

 The controlled validation cases examine whether adjoint-derived priorities are consistent with the stress mechanisms governing secondary motion and separation, and whether the resulting weighting improves deployed predictions. For each case, the analysis connects the global component weights in Table~\ref{tab:global-adjoint-weights} to anisotropy reconstruction and the subsequent velocity response.

\begin{table}[htbp]
    \centering
    \small
    \renewcommand{\arraystretch}{1.15}
    \begin{tabular}{@{}lrrrrrr@{}}
        \toprule
        Configuration &
        \(W_{xx}\) &
        \(W_{xy}\) &
        \(W_{xz}\) &
        \(W_{yy}\) &
        \(W_{yz}\) &
        \(W_{zz}\) \\
        \midrule
        Square duct (\(U_y\)) &
        0.133 & 0.131 & 0.131 &
        \textbf{1.49} & \textbf{2.63} & \textbf{1.49} \\

        Periodic hill (\(U_x\)) &
        0.478 & \textbf{4.39} & 0.219 &
        0.477 & 0.219 & 0.221 \\
        \bottomrule
    \end{tabular}
    \caption{Global adjoint-derived component weights used to train the
    square-duct and periodic-hill closures. The weights are normalised to
    have a componentwise mean of one; bold values identify the prioritised components in each configuration.}
    \label{tab:global-adjoint-weights}
\end{table}

\subsubsection{Secondary flow in the square duct}
\label{sec:results-duct}

The square-duct case tests whether goal-oriented weighting identifies the stress components relevant to secondary motion and improves its prediction. The adjoint-derived weights prioritise the cross-plane anisotropy components \(b_{yy}\), \(b_{zz}\), and \(b_{yz}\) for the selected \(U_y\) objective (Table~\ref{tab:global-adjoint-weights}). This prioritisation is consistent with the streamwise-vorticity balance, in which spatial derivatives of both the normal-stress difference \(\tau_{yy}-\tau_{zz}\) and the cross-plane shear stress \(\tau_{yz}\) contribute to secondary motion \citep{vidal2018turbulent,launder2002closure}. The baseline \(k\)--\(\omega\) model produces essentially no cross-plane normal-stress imbalance and fails to recover the observed secondary motion.

\begin{figure}[htbp]
    \centering
    \includegraphics[width=0.9\textwidth]{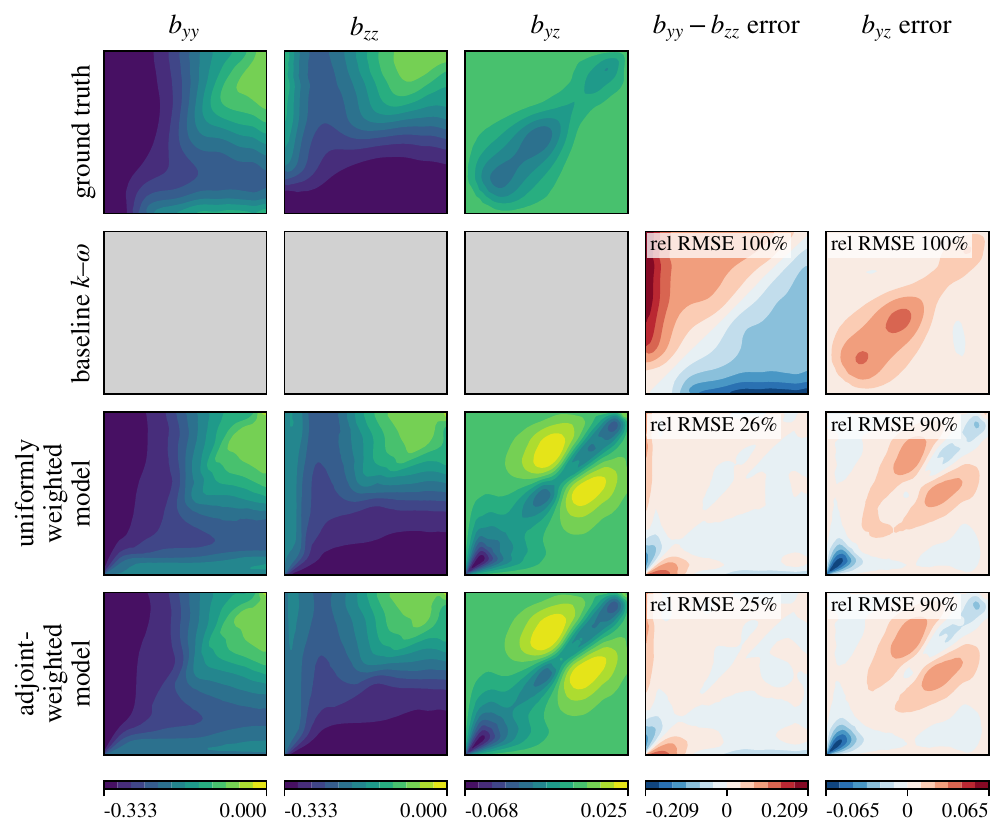}
    \caption{\emph{A priori} comparison of the cross-plane anisotropy components relevant to secondary motion in the square duct at \(\mathrm{Re}_{\mathrm{b}}=2600\). Rows compare the DNS reference, baseline \(k\)--\(\omega\) model, uniformly weighted model, and adjoint-weighted model. The first three columns show \(b_{yy}\), \(b_{zz}\), and \(b_{yz}\); the final two show the corresponding model errors in \(b_{yy}-b_{zz}\) and \(b_{yz}\), with relative RMSE reported in each error panel. A common colour scale is used within each column}
    \label{fig:duct-apriori}
\end{figure}

Both learned models recover cross-plane anisotropy absent from the baseline prediction, while adjoint weighting gives a modest improvement in the normal-stress imbalance. The uniformly weighted model reconstructs the principal spatial structure of \(b_{yy}-b_{zz}\), and the adjoint-weighted model further reduces its reconstruction error (Fig.~\ref{fig:duct-apriori}). The reconstruction of \(b_{yz}\) remains similar for the two learned models. 

\begin{figure}[htbp]
    \centering
    \includegraphics[width=0.9\textwidth]{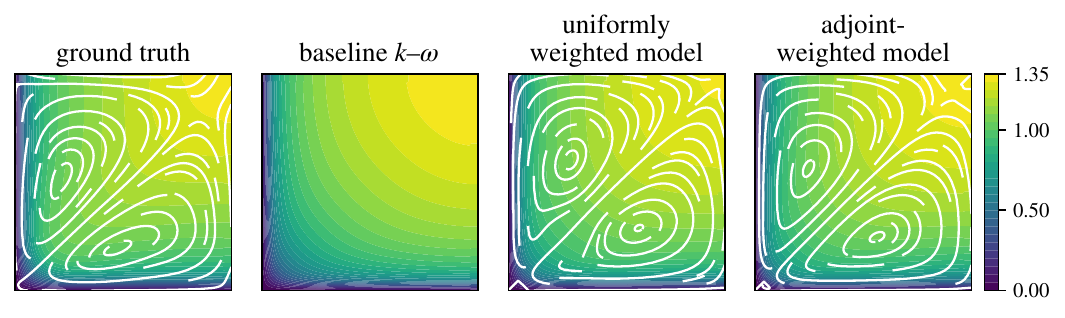}
    \caption{Recovery of secondary motion in the square duct at
    \(\mathrm{Re}_{\mathrm{b}}=2600\). Normalised streamwise-velocity contours are overlaid with streamlines of the cross-plane velocity \((U_y,U_z)\). From left to right, the panels show the DNS reference, the baseline \(k\)--\(\omega\) model, the uniformly weighted model, and the adjoint-weighted model}
    \label{fig:duct-secondary-flow}
\end{figure}

\begin{figure}[htbp]
    \centering
    \includegraphics[width=0.6\linewidth]{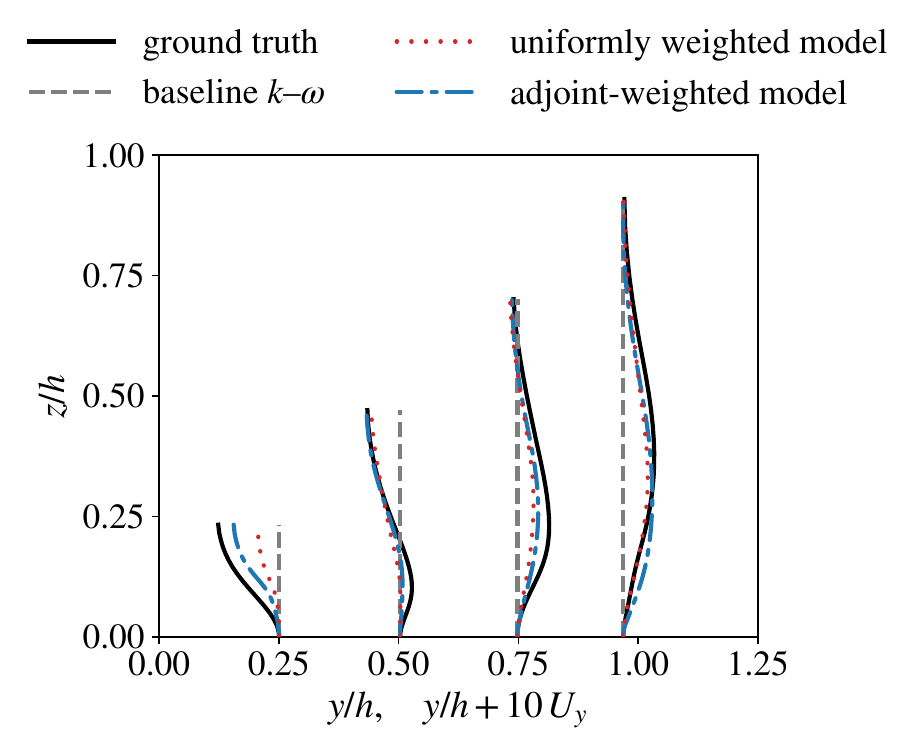}
    \caption{Quantitative comparison of the in-plane velocity in the square duct at \(\mathrm{Re}_{\mathrm{b}}=2600\). Profiles of \(U_y\) along \(z/h\) are
    sampled at \(y/h=0.25\), \(0.50\), \(0.75\), and \(1.00\), and are
    displayed using the offset \(y/h+10U_y\). Predictions from the baseline \(k\)--\(\omega\) model, the uniformly weighted model, and the adjoint-weighted model are compared with the DNS reference}
    \label{fig:duct-Uy-profiles}
\end{figure}

The adjoint-weighted model gives a more accurate secondary-flow prediction at the training condition \(\mathrm{Re}_{\mathrm{b}}=2600\). Both learned models recover the characteristic cross-plane vortices, whereas the baseline \(k\)--\(\omega\) model predicts essentially zero in-plane velocity (Fig.~\ref{fig:duct-secondary-flow}). The adjoint-weighted model follows the DNS \(U_y\) profiles more closely across the sampled locations (Fig.~\ref{fig:duct-Uy-profiles}), consistent with its reduced reconstruction error in the normal-stress imbalance.


\begin{figure}[htbp]
    \centering
\includegraphics[width=0.65\textwidth]{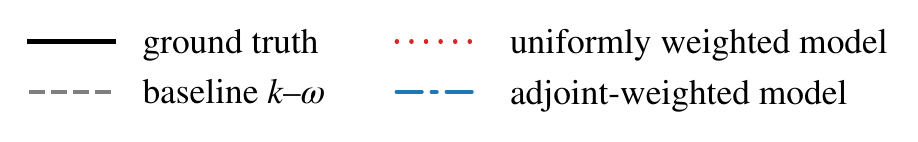}
    \vspace{0.5em}
    \begin{subfigure}[t]{0.48\textwidth}
        \centering
        \includegraphics[width=\linewidth]{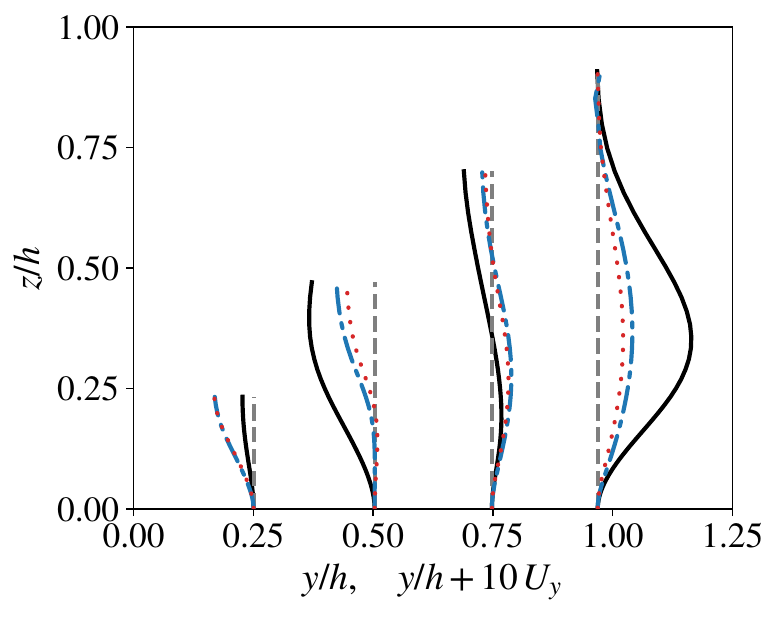}
        \caption{\(\mathrm{Re}_{\mathrm{b}}=1100\)}
        \label{fig:duct-transfer-1100}
    \end{subfigure}
    \hfill
    \begin{subfigure}[t]{0.48\textwidth}
        \centering
        \includegraphics[width=\linewidth]{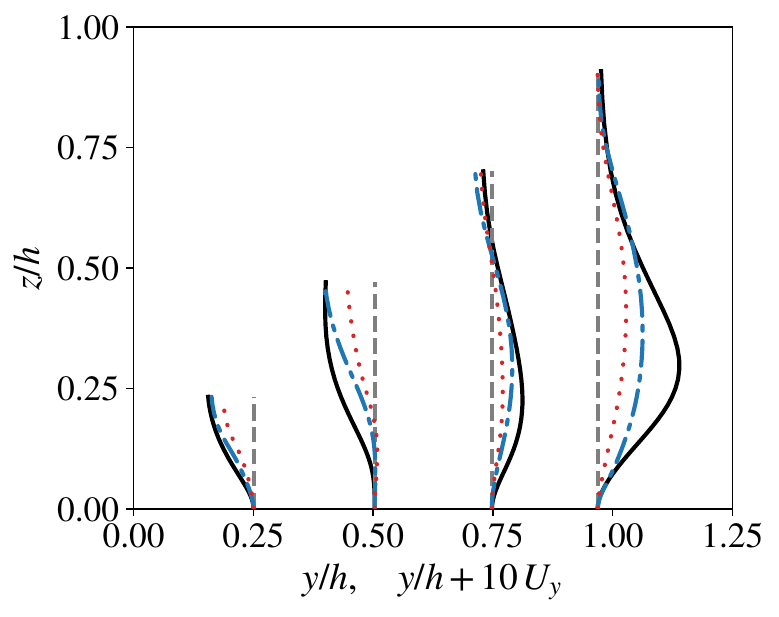}
        \caption{\(\mathrm{Re}_{\mathrm{b}}=1800\)}
        \label{fig:duct-transfer-1800}
    \end{subfigure}

    \caption{Generalisation of the square-duct closures across Reynolds numbers. In-plane velocity profiles are compared with DNS at the test conditions (a) \(\mathrm{Re}_{\mathrm{b}}=1100\) and (b) \(\mathrm{Re}_{\mathrm{b}}=1800\). Successive profiles are displayed using \(y/h+10U_y\). Both learned models are trained at \(\mathrm{Re}_{\mathrm{b}}=2600\), and the adjoint weights are constructed only from that training condition}
    \label{fig:duct-transfer}
\end{figure}

Generalisation of the square-duct closure is assessed at the test conditions \(\mathrm{Re}_{\mathrm{b}}=1100\) and \(1800\). The adjoint-weighted closure gives closer overall agreement with the DNS secondary-flow profiles at both Reynolds numbers (Fig.~\ref{fig:duct-transfer}). These results support the transfer of the weighting benefit beyond the training condition, with the closure and weights retained from \(\mathrm{Re}_{\mathrm{b}}=2600\).

\FloatBarrier
\subsubsection{Separated flow over the periodic hill}
\label{sec:results-hill}

\begin{figure}[htbp]
    \centering
    \includegraphics[width=1\textwidth]{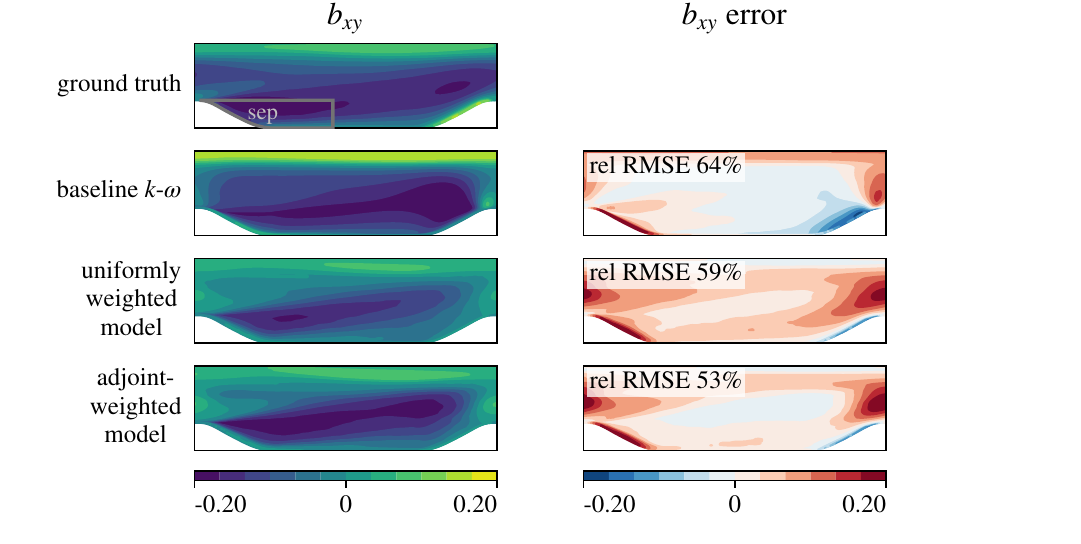}
    \caption{\emph{A priori} representation of the shear anisotropy \(b_{xy}\) for the periodic hill at the training geometry \(\alpha=1.5\). The DNS reference, baseline \(k\)--\(\omega\) model, uniformly weighted model, and adjoint-weighted model are compared using a common colour scale. The corresponding error fields and relative root-mean-square errors are reported for the three model predictions. The grey outline labelled \emph{sep} in the DNS panel delimits the separated region, $0<x/h<5$ and $0<y/h<1$, where \(h\) denotes the reference hill height, over which the separated-region velocity errors of Table~\ref{tab:hill-posteriori-errors} are evaluated; the region follows the definition of \citet{zhang2022ensemble}}
    \label{fig:hill-apriori-bxy}
\end{figure}

The periodic-hill case tests whether prioritising the shear anisotropy \(b_{xy}\) improves the prediction of separated-flow development. The adverse pressure gradient produces separation and a free shear layer, across which the Reynolds shear stress associated with \(b_{xy}\) transports streamwise momentum toward the recirculation region \citep{breuer2009flow}. Consistent with this mechanism, the \(U_x\)-based adjoint assigns the greatest component priority to \(b_{xy}\). The uniformly weighted model improves its spatial representation relative to the baseline \(k\)--\(\omega\) model, while the adjoint-weighted model further reduces the reconstruction error and gives the closest agreement with DNS at the training geometry \(\alpha=1.5\) (Fig.~\ref{fig:hill-apriori-bxy}).

\begin{figure}[htbp]
    \centering
    
    \includegraphics[width=0.65\textwidth]{figures/Fig8_10_11_legend.pdf}
    \vspace{0.5em}
    
    \begin{subfigure}[t]{0.49\textwidth}
        \centering
        \includegraphics[width=\linewidth]{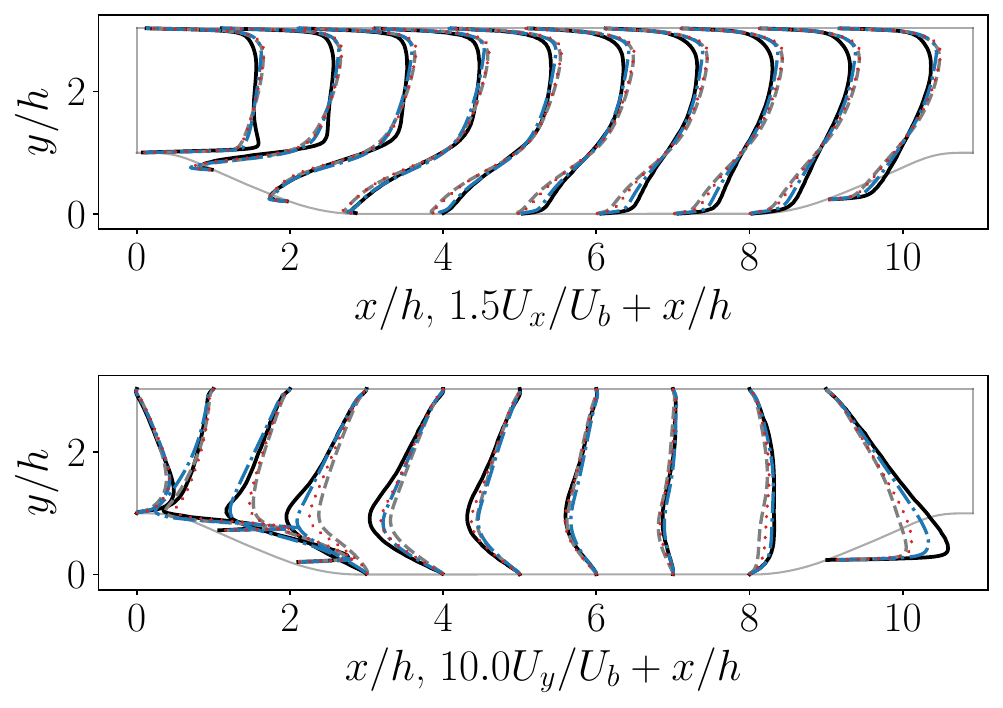}
        \caption{Mean-velocity profiles}
        \label{fig:hill-training-velocity}
    \end{subfigure}
    \hfill
    \begin{subfigure}[t]{0.49\textwidth}
        \centering
        \includegraphics[width=\linewidth]{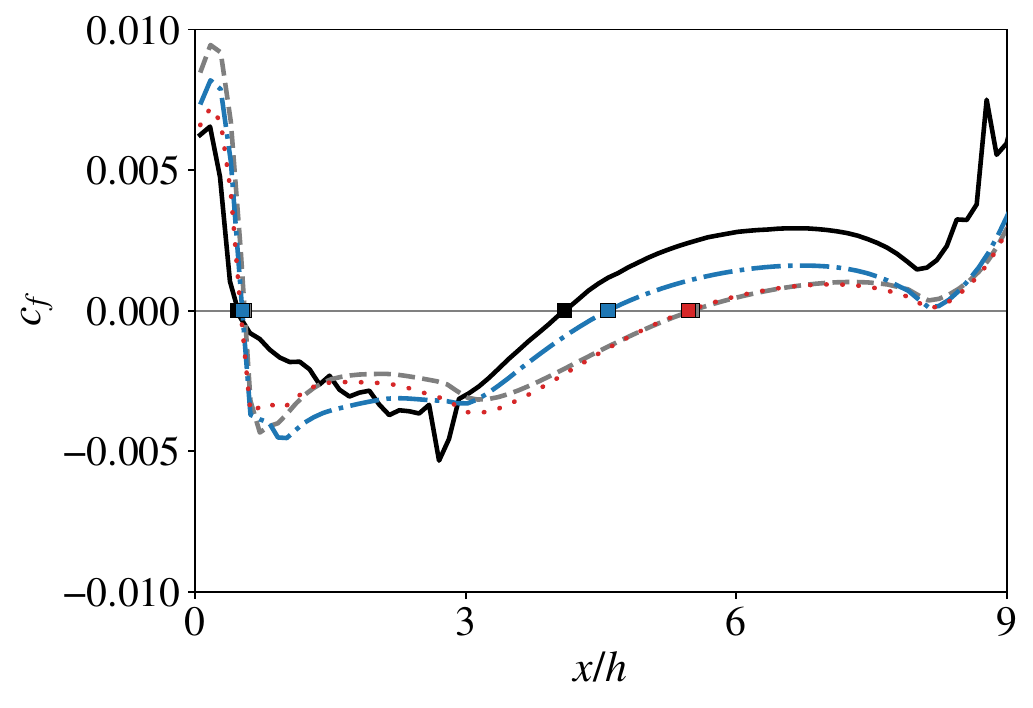}
        \caption{Lower-wall skin-friction coefficient}
        \label{fig:hill-training-cf}
    \end{subfigure}

    \caption{\emph{A posteriori} separated-flow prediction for the periodic hill
    at the training geometry \(\alpha=1.5\). Panel (a) compares the
    streamwise and wall-normal velocity profiles at the sampled
    streamwise locations. Successive profiles are offset according to the
    scaling indicated on each horizontal axis. Panel (b) compares the
    lower-wall skin-friction coefficient \(c_f\); the markers identify its
    zero crossings.}
    \label{fig:hill-training}
\end{figure}

After deployment at the training geometry \(\alpha=1.5\), the improved representation of \(b_{xy}\) is accompanied by a more accurate velocity prediction. The uniformly weighted model yields only modest changes relative to the baseline, whereas the adjoint-weighted model follows the DNS velocity profiles more closely (Fig.~\ref{fig:hill-training}(a)). Velocity errors are evaluated over both the full domain and a fixed separation-region window, marked in Fig.~\ref{fig:hill-apriori-bxy}. The full-field relative errors in \(U_x\) and \(U_y\) decrease to \(5.3\%\) and \(13.7\%\), respectively, for the adjoint-weighted model. Within the separation-region window, the corresponding errors decrease from \(22.1\%\) and \(31.7\%\) for the uniformly weighted model to \(12.1\%\) and \(19.2\%\) (Table~\ref{tab:hill-posteriori-errors}). 

\begin{table}[htbp]
    \centering
    \small
    \renewcommand{\arraystretch}{1.15}
    \begin{tabular}{@{}lcccc@{}}
        \toprule
        &
        \multicolumn{2}{c}{Full field} &
        \multicolumn{2}{c}{Separated region} \\
        \cmidrule(lr){2-3}\cmidrule(lr){4-5}
        Model &
        \(U_x\) error (\%) & \(U_y\) error (\%) &
        \(U_x\) error (\%) & \(U_y\) error (\%) \\
        \midrule
        Baseline \(k\)--\(\omega\) & 9.4 & 25.9 & 24.3 & 39.3 \\
        Uniformly weighted model  & 8.8 & 21.1 & 22.1 & 31.7 \\
        Adjoint-weighted model    & \textbf{5.3} & \textbf{13.7} & \textbf{12.1} & \textbf{19.2} \\
        \bottomrule
    \end{tabular}
    \caption{Volume-weighted relative root-mean-square errors in \emph{a posteriori}
      velocity predictions for the periodic hill at the training geometry
      \(\alpha=1.5\), evaluated over the full field and over the separated
      region marked in
      Fig.~\ref{fig:hill-apriori-bxy}. For each velocity component, the relative error is the volume-weighted norm of the prediction error divided by the corresponding reference-velocity norm over the same evaluation region, expressed as a percentage.}
    \label{tab:hill-posteriori-errors}
\end{table}

The velocity improvement is accompanied by a more accurate reattachment prediction. The skin-friction coefficient is defined as \(c_f=2\tau_w/U_b^2\), where \(\tau_w\) is the signed kinematic wall shear stress and \(U_b\) is the bulk reference velocity. The downstream zero crossing of \(c_f\) from the adjoint-weighted model moves toward the DNS location, whereas the baseline and uniformly weighted models retain a longer separated region (Fig.~\ref{fig:hill-training}(b)).

\begin{figure}[htbp]
    \centering

    \includegraphics[width=0.65\textwidth]{figures/Fig8_10_11_legend.pdf}
    \vspace{0.5em}

    \begin{subfigure}[t]{0.49\textwidth}
        \centering
        \includegraphics[width=\linewidth]{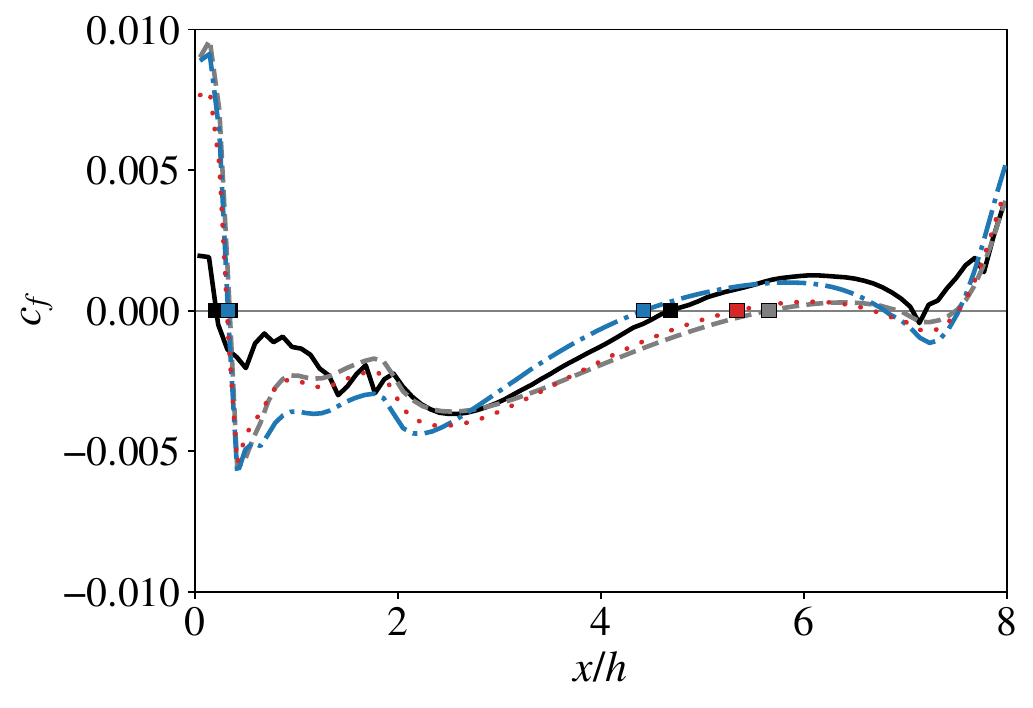}
        \caption{\(\alpha=1.0\)}
        \label{fig:hill-transfer-alpha10}
    \end{subfigure}
    \hfill
    \begin{subfigure}[t]{0.49\textwidth}
        \centering
        \includegraphics[width=\linewidth]{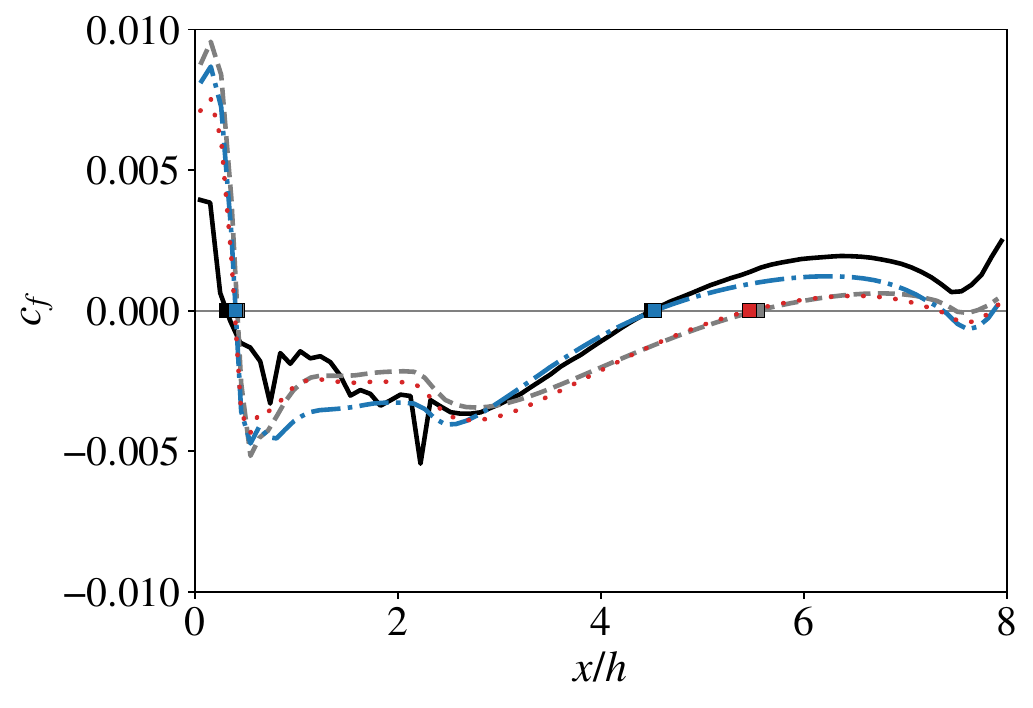}
        \caption{\(\alpha=1.2\)}
        \label{fig:hill-transfer-alpha12}
    \end{subfigure}

    \caption{Generalisation of the periodic-hill closures across geometry variations.
    Lower-wall skin-friction distributions are compared at the test geometries (a) \(\alpha=1.0\) and (b) \(\alpha=1.2\). Both learned models are trained at \(\alpha=1.5\), and the adjoint weights are constructed only
    from that training geometry. Markers indicate the zero crossings used
    to identify separation and reattachment}
    \label{fig:hill-transfer}
\end{figure}

The improvement in reattachment prediction persists at the test geometries \(\alpha=1.0\) and \(1.2\). At both geometries, the adjoint-weighted model predicts the downstream zero crossing of \(c_f\) closer to DNS than the baseline and uniformly weighted models (Fig.~\ref{fig:hill-transfer}). The closure and weights obtained at \(\alpha=1.5\) are retained, demonstrating that the benefit extends beyond the training geometry.

\FloatBarrier
\subsection{Jet-in-crossflow}
\label{sec:results-jic}
The jet-in-crossflow results distinguish improvements in the internal hole flow from improvements in downstream velocity and cooling predictions. Both learned closures improve the in-hole separation, but the adjoint-weighted model also reduces the downstream discrepancies retained by uniformly weighted training. The following analysis connects this comparison to the spatially varying training priorities, examines the resulting coolant distribution, and evaluates cooling effectiveness at the test blowing ratio.

\subsubsection{Spatial and component priorities}

\begin{figure}[htbp]
    \centering

    \begin{subfigure}[t]{1\textwidth}
        \centering
        \includegraphics[
            width=0.95\linewidth,
            keepaspectratio
        ]{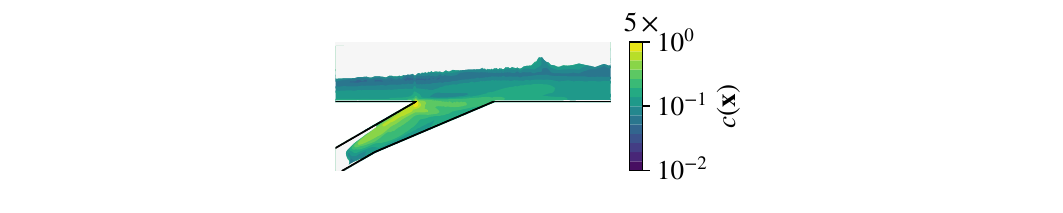}
        \caption{Cell-relevance factor \(c(\mathbf{x})\)}
        \label{fig:jic-cell-magnitude}
    \end{subfigure}

    \begin{subfigure}[t]{1\textwidth}
        \centering
        \includegraphics[
            width=0.95\linewidth,
keepaspectratio]{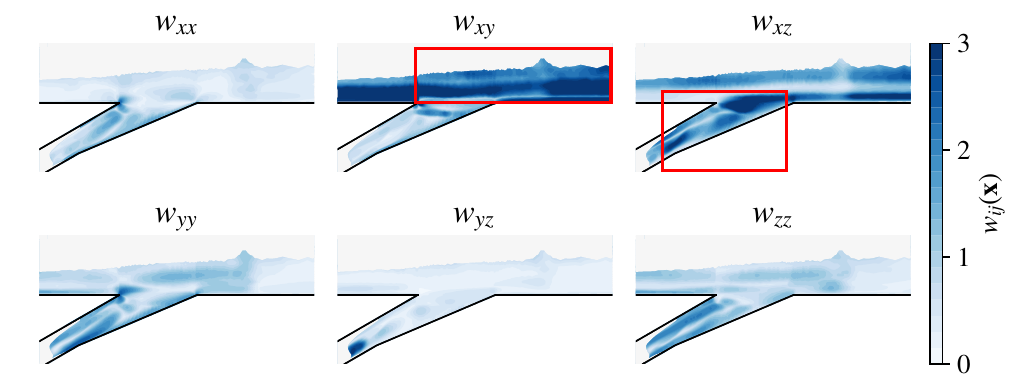}
        \caption{Normalised component factors
        \(w_{ij}(\mathbf{x})\)}
        \label{fig:jic-component-weights}
    \end{subfigure}

    \caption{Spatial and componentwise distribution of the adjoint-derived training weights for the jet-in-crossflow condition \(\mathrm{BR}=1.5\), obtained from the baseline-RANS \(U_x\) adjoint. Panel (a) shows the cell-relevance factor \(c(\mathbf{x})\), and panel (b) shows the normalised component factors \(w_{ij}(\mathbf{x})\). Coloured fields represent weights in the retained training region; light-grey areas indicate cells excluded by the mask, and black lines mark the displayed physical boundaries. Red rectangles highlight regions of comparatively large \(w_{xy}\) and \(w_{xz}\)}
    \label{fig:jic-cellwise-weights}
\end{figure}

The adjoint-derived weights show that the Reynolds-stress information relevant to the downstream \(U_x\) objective varies in both magnitude and component across the jet-in-crossflow domain. The cell-relevance factor \(c(\mathbf{x})\) concentrates the fitting emphasis from the diffusing hole through the hole-exit region and into the downstream near-wall jet, while cells outside the adjoint mask have comparatively small sensitivity in the local linearisation about the baseline RANS state. Within the retained region, the component priorities also change spatially: \(b_{xy}\) receives the greatest emphasis in the downstream shear layer, whereas the relative importance of \(b_{xz}\) increases inside the hole (Fig.~\ref{fig:jic-cellwise-weights}). A spatially aggregated component weight would average these distinct patterns, whereas the cellwise formulation preserves the local stress information associated with the selected velocity response.

The effectiveness of these spatially varying priorities is assessed through the deployed flow predictions. A domain-averaged anisotropy error measures overall reconstruction accuracy, but does not by itself establish whether the selected velocity prediction improves.

\subsubsection{Velocity development}

\begin{figure}[htbp]
    \centering
    \includegraphics[width=0.75\textwidth]{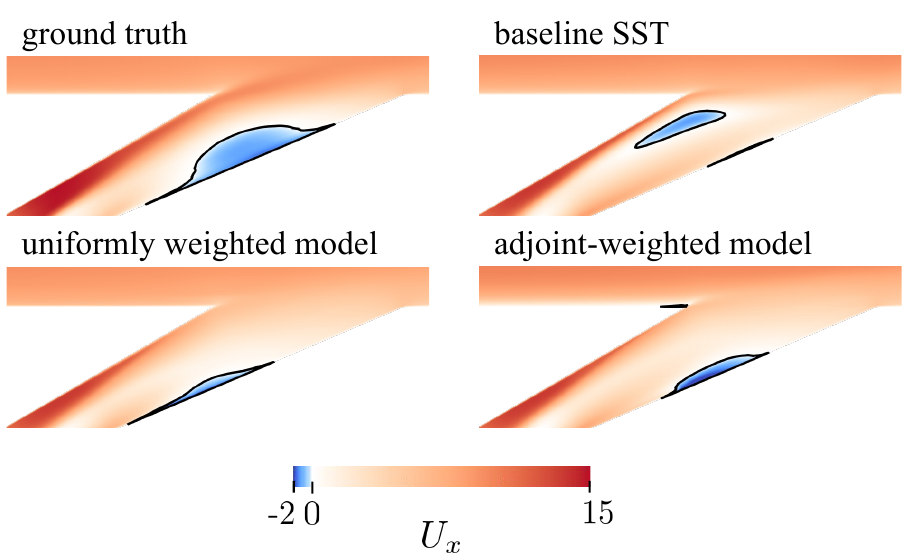}
    \caption{Streamwise-velocity distribution used to assess the separation inside the diffusing hole and the downstream near-wall flow at the jet-in-crossflow training condition \(\mathrm{BR}=1.5\). Centre-plane predictions from the baseline SST, uniformly weighted, and adjoint-weighted models are compared with the LES reference using common contour limits. Blue regions indicate negative \(U_x\), with black zero-velocity contours delineating reversed flow}
    \label{fig:jic-Ux-contours}
\end{figure}

The baseline \(k\)--\(\omega\) SST model misplaces the separation inside the diffusing hole and distorts the \(U_x\) distribution approaching the exit (Fig.~\ref{fig:jic-Ux-contours}). Related deficiencies in diffused film-cooling holes have been associated with distorted exit flow and downstream coolant distributions \citep{Zhang2023DoubleExpansion,Schroeder2014Baseline,Gunady2021Velocity}.

\begin{figure}[htbp]
    \centering
    \includegraphics[width=0.75\textwidth]{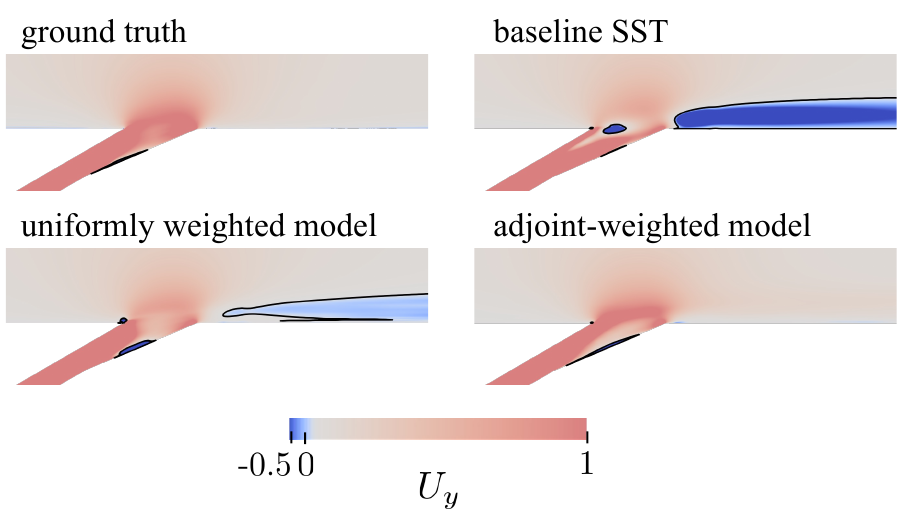}
    \caption{Wall-normal-velocity distribution used to assess the vertical development of the injected jet at the jet-in-crossflow training condition \(\mathrm{BR}=1.5\). Centre-plane predictions from the baseline SST, uniformly weighted, and adjoint-weighted models are compared with the LES reference using common contour limits. Blue regions indicate negative \(U_y\), corresponding to flow directed toward the wall; black contours mark zero wall-normal velocity}
    \label{fig:jic-Uy-contours}
\end{figure}

Both learned models improve the in-hole separation, but the adjoint-weighted model gives closer agreement with LES downstream of the exit. The uniformly weighted model retains a pronounced near-wall deficit in \(U_x\) and an extensive region of negative \(U_y\) (Figs.~\ref{fig:jic-Ux-contours} and \ref{fig:jic-Uy-contours}). Adjoint weighting reduces both discrepancies while preserving the improved internal separation. The benefit of weighting therefore extends beyond the correction of the in-hole flow.

\begin{figure}[htbp]
    \centering
    \includegraphics[width=0.7\textwidth]{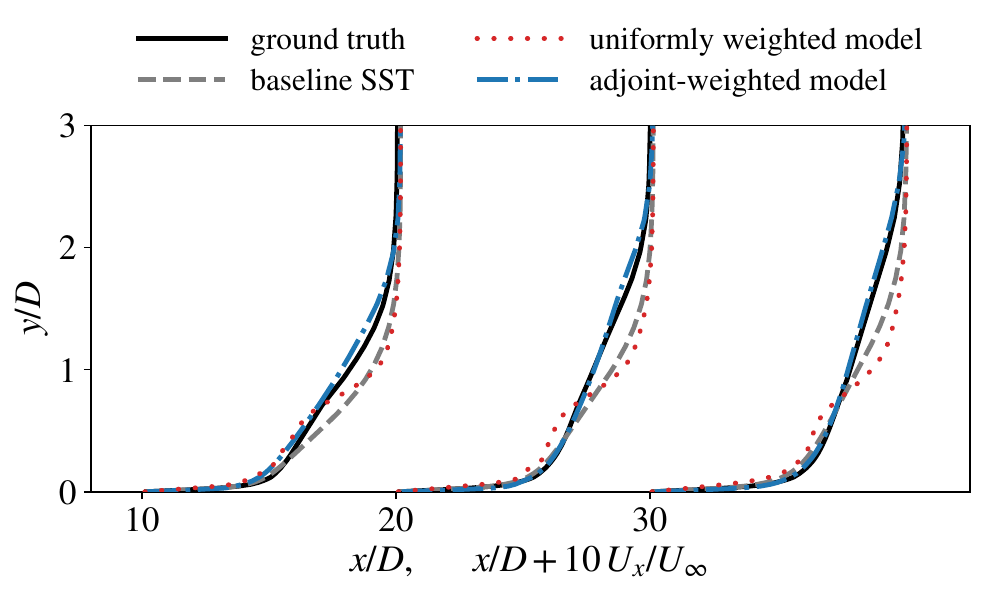}
    \caption{Downstream development of the centre-plane streamwise velocity at the jet-in-crossflow training condition \(\mathrm{BR}=1.5\). Profiles from the baseline SST, uniformly weighted, and adjoint-weighted models are compared with the LES reference at \(x/D=5,10,15,20,25,30,\) and \(35\). Successive profiles are displayed using the offset \(x/D+10U_x/U_\infty\)}
    \label{fig:jic-Ux-profiles}
\end{figure}

The downstream profiles locate the remaining streamwise-velocity errors of the uniformly weighted model in two distinct regions. Across most sampled stations, the uniformly weighted model underpredicts \(U_x\) in the near-wall region \(y/D\lesssim1\) and overpredicts it in the upper jet shear layer (Fig.~\ref{fig:jic-Ux-profiles}). The adjoint-weighted model is closer to LES in both regions, while differences among the predictions diminish toward the outer flow. These profiles provide a quantitative comparison of the velocity component used to construct the adjoint weights.

\subsubsection{Coolant distribution and cooling effectiveness}

\begin{figure}[htbp]
    \centering
    \includegraphics[width=0.75\textwidth]{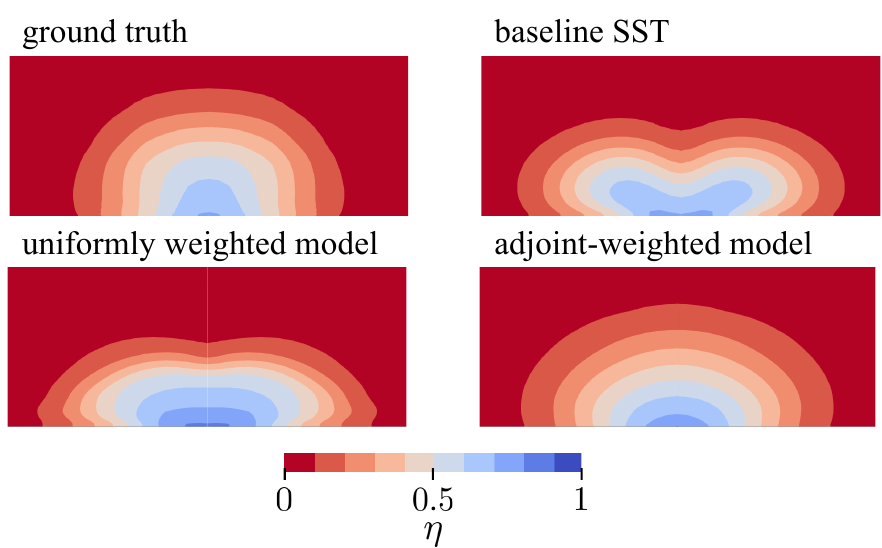}
    \caption{Cross-sectional coolant distribution at \(x/D=10\) for the jet-in-crossflow training condition \(\mathrm{BR}=1.5\). Normalised-temperature predictions \(\eta\) from the baseline SST, uniformly weighted, and adjoint-weighted models are compared with the LES reference using common contour limits}
    \label{fig:jic-temperature-x10}
\end{figure}

The improved downstream velocity prediction is accompanied by a more centred coolant distribution at \(x/D=10\), closer to the LES reference. The adjoint-weighted model recovers a more centred cross-sectional coolant distribution at \(x/D=10\). The baseline SST and uniformly weighted models predict a vertically compressed plume with two off-centre peaks, while the adjoint-weighted prediction is closer to the central structure observed in LES (Fig.~\ref{fig:jic-temperature-x10}). Similar bimodal distributions have been associated with diffuser separation and a persistent central velocity deficit in film-cooling flows \citep{Haydt2018AreaRatio}.

\begin{figure}[htbp]
    \centering
    \includegraphics[width=0.95\textwidth]{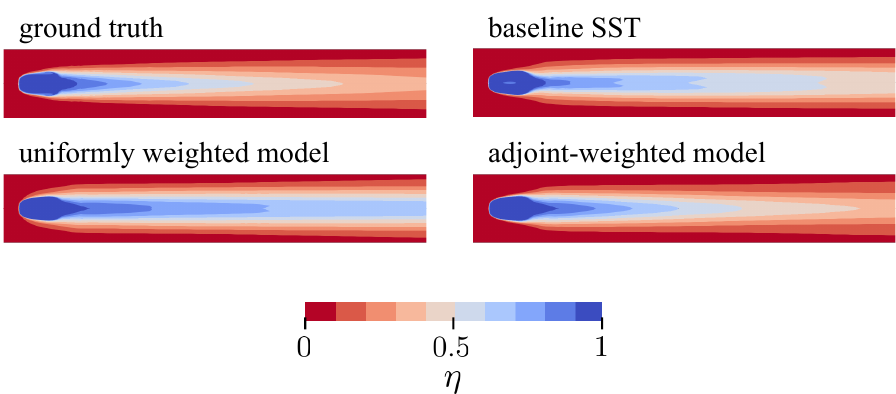}
    \caption{Wall cooling-effectiveness distribution for the jet-in-crossflow training condition \(\mathrm{BR}=1.5\). Predictions from the baseline SST, uniformly weighted, and adjoint-weighted models are compared with the LES reference using common contour limits, allowing the lateral position and downstream development of the coolant footprint to be assessed}
    \label{fig:jic-eta-wall}
\end{figure}

The adjoint-weighted model also recovers a more centred wall-cooling footprint, closer to LES. The uniformly weighted prediction retains two laterally displaced regions of high cooling effectiveness (Fig.~\ref{fig:jic-eta-wall}). The split footprint persists despite the uniformly weighted model's improved in-hole separation and is accompanied by remaining downstream velocity discrepancies.

\begin{figure}[htbp]
    \centering
    
    \includegraphics[width=0.65\textwidth]{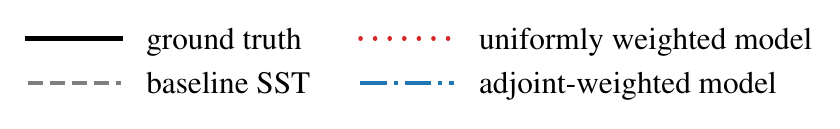}
    \vspace{0.5em}
    
    \begin{subfigure}[t]{0.48\textwidth}
        \centering
        \includegraphics[width=\linewidth]{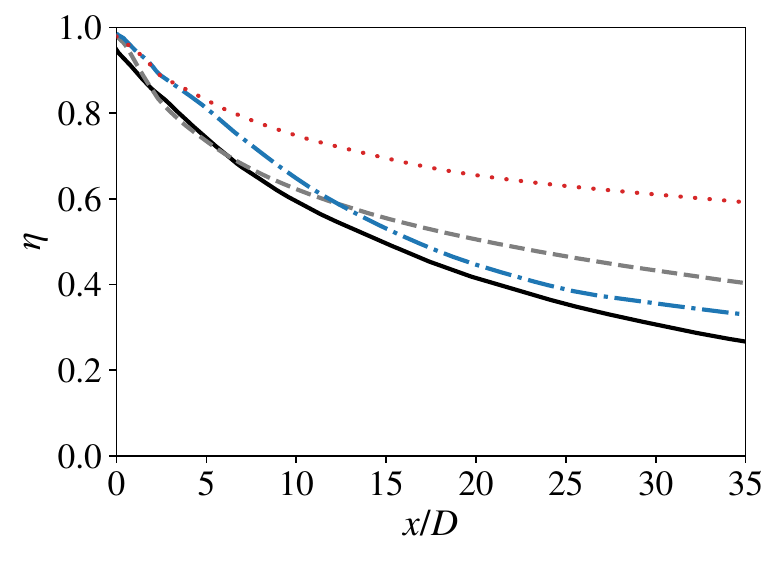}
        \caption{Training condition, \(\mathrm{BR}=1.5\)}
        \label{fig:jic-eta-br15}
    \end{subfigure}
    \hfill
    \begin{subfigure}[t]{0.48\textwidth}
        \centering
        \includegraphics[width=\linewidth]{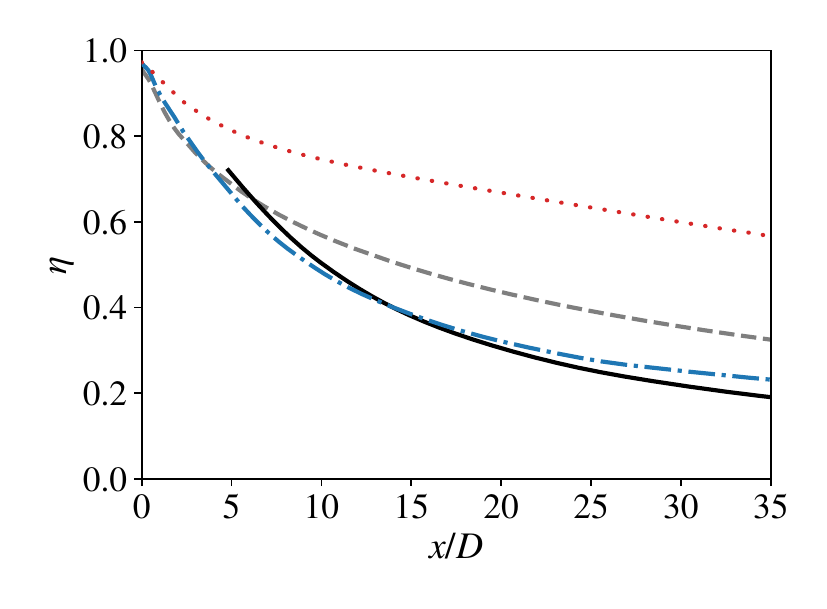}
        \caption{Test condition, \(\mathrm{BR}=1.0\)}
        \label{fig:jic-eta-br10}
    \end{subfigure}

    \caption{Centreline wall cooling effectiveness used to assess the downstream scalar response at the training condition and its generalisation to an unseen blowing ratio. High-fidelity reference data (LES data for \(\mathrm{BR}=1.5\) and experimental data for \(\mathrm{BR}=1.0\)) are compared with predictions from the baseline SST model, the uniformly weighted model, and the adjoint-weighted model}
    \label{fig:jic-eta-lines}
\end{figure}

At the training condition \(\mathrm{BR}=1.5\), adjoint weighting reduces the centreline cooling-effectiveness overprediction produced by uniformly weighted training. The uniformly weighted model deviates more strongly from LES than the baseline SST model over most of the downstream region, whereas the adjoint-weighted prediction follows the LES decay more closely, particularly farther downstream (Fig.~\ref{fig:jic-eta-lines}(a)).

The cooling-effectiveness benefit also persists at the test condition \(\mathrm{BR}=1.0\). Without retraining the closure or recomputing its weights, the adjoint-weighted model agrees more closely with the experimental reference \citep{Schroeder2014Baseline}, while the uniformly weighted model retains a substantial downstream overprediction (Fig.~\ref{fig:jic-eta-lines}(b)). This comparison supports transfer of the cooling-effectiveness prediction between the two tested operating conditions; velocity-field accuracy at the test condition is not independently assessed.

The thermal improvements are obtained without temperature or heat-flux information in the adjoint objective. They demonstrate a beneficial response of the coupled momentum--scalar system to the weighted turbulence closure, but do not establish an improvement in the turbulent heat-flux model itself.

\FloatBarrier
\section{Conclusion}

In this work we propose a goal-oriented direct-training framework that uses a single offline adjoint solution to weight high-fidelity Reynolds-stress data according to their influence on a selected velocity QoI. The resulting weights account for differences in relevance across anisotropy components and spatial locations, while subsequent model optimisation remains solver-free. The framework therefore introduces information about the deployed flow response into direct turbulence-model training without requiring repeated RANS evaluations during parameter optimisation.

The square-duct and periodic-hill cases provide controlled validation of the weighting principle, while the jet in crossflow demonstrates its application to a complex flow with spatially varying component relevance. In the square duct, adjoint-derived weighting reduces the error in the Reynolds normal-stress imbalance governing secondary motion and improves the in-plane velocity prediction. In the periodic-hill flow, weighting emphasises the shear anisotropy responsible for momentum transport across the separated shear layer, improving the velocity and reattachment predictions at both the training and test geometries. The jet in crossflow provides the principal demonstration of cellwise weighting. Although the uniformly weighted model improves the gross separation inside the diffusing hole, it retains inaccurate downstream velocity development and substantially overpredicts the cooling effectiveness. The adjoint-weighted model preserves the improved in-hole separation while producing downstream velocity, coolant-distribution, and cooling-effectiveness predictions closer to the reference data. The improved cooling-effectiveness prediction at the test condition \(\mathrm{BR}=1.0\) provides evidence of generalisation between the two operating conditions considered for the same jet-in-crossflow configuration. Collectively, these results show that reducing a uniformly aggregated constitutive error does not guarantee an improved \emph{a posteriori} prediction; weighting the constitutive data according to their influence on the intended flow quantity provides a more targeted training loss for direct turbulence-model training.

The present results also define the scope of the framework. The weights are obtained from a local adjoint sensitivity about the baseline RANS state and therefore do not capture higher-order responses to finite closure changes. In the jet-in-crossflow case, the adjoint objective acts only through the momentum equations, so the improvements in temperature and cooling effectiveness are downstream responses of the coupled momentum--scalar system and were not directly optimised. Future work should examine iteratively updated sensitivities for large closure changes, extend the adjoint formulation to coupled momentum and scalar objectives, and evaluate the framework across a broader range of complex flows and QoIs. More broadly, this work suggests that supervised learning can benefit from prioritising training data according to their influence on the final prediction, using offline sensitivity information to retain efficient training.

\begin{appendices}

\section{Tensor-basis neural network representation}
\label{app:tbnn-details}

\setcounter{table}{3}
\renewcommand{\thetable}{\arabic{table}}

This appendix specifies the tensor bases, scalar inputs, and normalisation used for the TBNN representation in Eq.~\eqref{eq:present-tbnn}. The scalar inputs are invariant under coordinate rotation, while the tensor bases transform consistently with the coordinate frame, yielding a frame-consistent prediction of the anisotropy tensor \citep{pope1975more,ling2016reynolds}.

The tensor bases are constructed from the mean strain- and rotation-rate tensors,
\begin{equation}
\mathbf{S}^{\mathrm{d}}
=
\operatorname{dev}\!\left[
\frac{1}{2}
\left(
\nabla\mathbf{U}
+
(\nabla\mathbf{U})^{\mathsf{T}}
\right)
\right],
\qquad
\boldsymbol{\Omega}
=
\frac{1}{2}
\left(
\nabla\mathbf{U}
-
(\nabla\mathbf{U})^{\mathsf{T}}
\right).
\label{eq:strain-rotation}
\end{equation}
Their dimensionless forms are
\begin{equation}
\tilde{\mathbf{S}}
=
\tau_s\mathbf{S}^{\mathrm{d}},
\qquad
\tilde{\boldsymbol{\Omega}}
=
\tau_s\boldsymbol{\Omega},
\qquad
\tau_s
=
\frac{1}{\beta^{*}\omega},
\qquad
\beta^{*}=0.09,
\label{eq:dimensionless-gradient-tensors}
\end{equation}
where \(\omega\) is the specific dissipation rate and \(\beta^{*}\) is the turbulent-kinetic-energy destruction coefficient in the \(k\)--\(\omega\) model. For construction of the training inputs, \(\omega\) is obtained by solving its modelled transport equation with the high-fidelity mean velocity and turbulent kinetic energy imposed.

The present quadratic representation retains the first four tensors of the general ten-term integrity basis:
\begin{equation}
\begin{aligned}
\boldsymbol{\mathcal{T}}^{(1)}
&=
\tilde{\mathbf{S}},
&
\boldsymbol{\mathcal{T}}^{(2)}
&=
\tilde{\mathbf{S}}\tilde{\boldsymbol{\Omega}}
-
\tilde{\boldsymbol{\Omega}}\tilde{\mathbf{S}},
\\
\boldsymbol{\mathcal{T}}^{(3)}
&=
\tilde{\mathbf{S}}^{\,2}
-
\frac{1}{3}
\operatorname{tr}\!\left(
\tilde{\mathbf{S}}^{\,2}
\right)\mathbf{I},
&
\boldsymbol{\mathcal{T}}^{(4)}
&=
\tilde{\boldsymbol{\Omega}}^{\,2}
-
\frac{1}{3}
\operatorname{tr}\!\left(
\tilde{\boldsymbol{\Omega}}^{\,2}
\right)\mathbf{I}.
\end{aligned}
\label{eq:retained-raw-bases}
\end{equation}
Following the tensor-basis normalisation strategy of \citet{ji2026enhancing}, the nonlinear bases are normalised to control variations in their magnitudes:
\begin{equation}
\mathbf{T}^{(1)}
=
\boldsymbol{\mathcal{T}}^{(1)},
\qquad
\mathbf{T}^{(n)}
=
\frac{\boldsymbol{\mathcal{T}}^{(n)}}
{\left\|\boldsymbol{\mathcal{T}}^{(n)}\right\|_F+\delta_T},
\quad
n=2,3,4,
\qquad
\delta_T=10^{-9}.
\label{eq:nonlinear-basis-normalization}
\end{equation}
Here, \(\|\cdot\|_F\) denotes the Frobenius norm, and \(\delta_T\) prevents division by zero. The linear basis remains unnormalised. The symbols \(\mathbf{T}^{(n)}\) in Eq.~\eqref{eq:present-tbnn} denote these processed bases.

The scalar inputs describe the local strain, rotation, viscosity ratio, and wall proximity. Bounded strain- and rotation-rate tensors are defined as
\begin{equation}
\widehat{\mathbf{S}}
=
\frac{\mathbf{S}^{\mathrm{d}}}
{\left\|\mathbf{S}^{\mathrm{d}}\right\|_F+\tau_s^{-1}},
\qquad
\widehat{\boldsymbol{\Omega}}
=
\frac{\boldsymbol{\Omega}}
{\left\|\boldsymbol{\Omega}\right\|_F+\tau_s^{-1}},
\label{eq:bounded-input-tensors}
\end{equation}
and are used only to construct the scalar inputs. The feature selection and normalisation follow \citet{liu2026toward}, with the definitions summarised in Table~\ref{tab:feature-definitions}, where \(d\) denotes the distance to the nearest wall.

\begin{table}[htbp]
\centering
\renewcommand{\arraystretch}{1.15}
\begin{tabular}{
m{0.12\linewidth}
>{\raggedright\arraybackslash}m{0.35\linewidth}
>{\raggedright\arraybackslash}m{0.42\linewidth}}
\toprule
\textbf{Feature} &
\textbf{Description} &
\textbf{Definition} \\
\midrule
\(q_1\) &
Strain-rate invariant &
\(\operatorname{tr}(\widehat{\mathbf{S}}^{\,2})\) \\
\(q_2\) &
Rotation-rate invariant &
\(\operatorname{tr}(\widehat{\boldsymbol{\Omega}}^{\,2})\) \\
\(q_3\) &
Third-order strain invariant &
\(\operatorname{tr}(\widehat{\mathbf{S}}^{\,3})\) \\
\(q_4\) &
Coupled strain--rotation invariant &
\(\operatorname{tr}(\widehat{\boldsymbol{\Omega}}^{\,2}\widehat{\mathbf{S}})\) \\
\(q_{\nu}\) &
Viscosity ratio &
\(\nu_t/(100\nu+\nu_t)\) \\
\(q_d\) &
Wall-distance Reynolds number &
\(\min\!\left(\sqrt{k}\,d/(50\nu),\,2\right)\) \\
\bottomrule
\end{tabular}
\caption{Invariant scalar inputs to the TBNN. The jet-in-crossflow model uses \(q_1\)--\(q_4\), \(q_{\nu}\), and \(q_d\). For the two-dimensional configurations, \(q_3\) and \(q_4\) provide no independent information and are therefore omitted}
\label{tab:feature-definitions}
\end{table}

The neural network maps the selected scalar inputs to the four coefficient functions \(g^{(n)}(\mathbf{q})\), which are combined with the processed bases through Eq.~\eqref{eq:present-tbnn} to obtain \(\mathbf{b}^{\mathrm{pred}}\).

\section{Adjoint sensitivity and weight construction}
\label{app:weights}

\setcounter{figure}{18}
\renewcommand{\thefigure}{\arabic{figure}}

This appendix specifies the adjoint formulation and sensitivity conversion supporting Section~\ref{sec:weighting}, together with the sensitivity-threshold selection and the construction of global component weights for the validation cases.

\textit{Adjoint formulation.} The adjoint equations evaluate the sensitivity of the velocity objective in Eq.~\eqref{eq:qoi} to Reynolds-stress perturbations about the baseline RANS state. With the Reynolds stress treated as the control variable, the representative incompressible formulation used here is
\begin{subequations}
\label{eq:adjoint-momentum}
\begin{align}
\nabla\cdot\hat{\mathbf{U}}&=0,
\label{eq:adjoint-continuity}\\
-(\mathbf{U}\cdot\nabla)\hat{\mathbf{U}}-\nu\nabla^2\hat{\mathbf{U}}+\nabla\hat{p}
&=-\frac{\partial J}{\partial\mathbf{U}},
\label{eq:adjoint-momentum-b}
\end{align}
\end{subequations}
where \(\mathbf{U}\) is the converged baseline velocity, and \(\hat{\mathbf{U}}\) and \(\hat{p}\) are the adjoint velocity and pressure. Following the stabilised formulation of \citet{michelenstrofer2021endtoend}, the transpose-convection term is omitted, giving an approximate local sensitivity. The jet-in-crossflow calculation uses the corresponding compressible RANS operator. Under the stress and adjoint sign conventions adopted here, the Reynolds-stress sensitivity is
\begin{equation}
\frac{\mathrm{d} J}{\mathrm{d}\boldsymbol{\tau}}=\nabla\hat{\mathbf{U}}.
\label{eq:stress-sensitivity}
\end{equation}

\textit{Anisotropy-component sensitivities.} The stress gradient is converted to the symmetric, traceless anisotropy representation while holding the baseline turbulent kinetic energy fixed. The relation \(\boldsymbol{\tau}=-\tfrac{2}{3}k\mathbf{I}-2k\mathbf{b}\) supplies the factor \(-2k\), giving the projected tensor gradient
\begin{equation}
\mathbf{H}=-2k\,\operatorname{dev}\!\left[\operatorname{sym}\!\left(\nabla\hat{\mathbf{U}}\right)\right],
\label{eq:anisotropy-sensitivity}
\end{equation}
where \(\operatorname{sym}(\mathbf{A})=(\mathbf{A}+\mathbf{A}^{\mathsf{T}})/2\). The symmetric and deviatoric projections enforce the symmetry and zero-trace constraints of the anisotropy tensor. The sensitivity magnitudes for the six unique components are
\begin{equation}
G_{ij}(\mathbf{x})=
\begin{cases}
|H_{ii}(\mathbf{x})|, & i=j,\\
2|H_{ij}(\mathbf{x})|, & i<j.
\end{cases}
\label{eq:component-sensitivity-conversion}
\end{equation}
Each unique off-diagonal component represents both \(b_{ij}\) and \(b_{ji}\), so the contributions of these two entries are combined through the factor of two. This factor is applied before spatial aggregation or weight normalisation. The diagonal entries retain the traceless projection; the six stored components are therefore not six independent degrees of freedom.

\begin{figure}[htbp]
    \centering
    \includegraphics[width=0.5\linewidth]{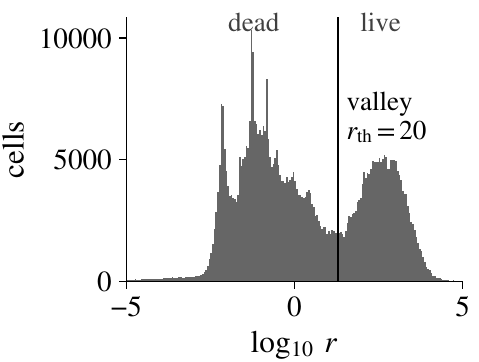}
    \caption{Selection of the sensitivity threshold for the jet-in-crossflow training region. The distribution of \(\log_{10}r\) contains low- and high-sensitivity modes, with \(r_{\mathrm{th}}=20\) selected at the intervening local minimum}
    \label{fig:jic-mask-distribution}
\end{figure}

\textit{Threshold selection.} For the jet-in-crossflow case, the normalised sensitivity score \(r\) defined in Eq.~\eqref{eq:cell-sensitivity-score} separates into low- and high-sensitivity modes when plotted as \(\log_{10}r\). The threshold \(r_{\mathrm{th}}=20\) is selected at the intervening local minimum (Fig.~\ref{fig:jic-mask-distribution}), giving the binary mask
\begin{equation}
M(\mathbf{x})=
\begin{cases}
1, & \mathbf{x}\in\Omega_{\mathrm{map}}\ \text{and}\ r(\mathbf{x})\ge r_{\mathrm{th}},\\
0, & \text{otherwise}.
\end{cases}
\label{eq:adjoint-mask}
\end{equation}

\textit{Global component weights.} For the square-duct and periodic-hill validation cases, each component sensitivity is averaged over the entire computational domain by summing the products of cell volume and \(G_{ij}\) and dividing by the total domain volume. All computational cells are included, without a spatial mask or threshold. Volume weighting prevents refined regions from receiving disproportionate influence solely because they contain more cells. These averages replace the local sensitivities in Eq.~\eqref{eq:cellwise-component-priority}, using the same transformation and parameter values. The resulting global weights \(W_{ij}\) have unit componentwise mean and are applied to all training cells.

The repeated weights in Table~\ref{tab:global-adjoint-weights} arise from two distinct mechanisms. The equal duct \(xy\) and \(xz\) weights and periodic-hill \(xz\) and \(yz\) weights result from negligible sensitivities and the common floor. In contrast, the nearly equal duct \(yy\) and \(zz\) weights and periodic-hill \(xx\) and \(yy\) weights reflect diagonal gradients of nearly equal magnitude and opposite sign, which yield nearly equal sensitivity magnitudes. These pairs are consistent with sensitivity to the corresponding normal-stress differences.

\end{appendices}

\section*{Acknowledgements}

The authors thank Christopher D. Ellis for insightful discussions on
jet-in-crossflow mechanisms and Jaeho Park for his support with the RANS simulations of the jet-in-crossflow. The authors
acknowledge the use of bwForCluster Helix, supported by the state of
Baden-W\"urttemberg through bwHPC and by the German Research Foundation
(DFG, grant INST 35/1597-1 FUGG). The LES computations were supported by the
National Supercomputing Center in South Korea with supercomputing resources
including technical support (Project Nos. TS-2026-RE-0009 and
KSC-2025-CRE-0028).

\section*{Statements and Declarations}

\noindent\textbf{Funding}

The authors acknowledge funding from the Deutsche Forschungsgemeinschaft
(DFG, German Research Foundation), project number 551388164, and from the
National Research Foundation of Korea (NRF) through a grant funded by the
Korea government (MSIT) (RS-2024-00438833). H.W. acknowledges further
funding by the Deutsche Forschungsgemeinschaft (DFG, German Research
Foundation) under Germany's Excellence Strategy -- EXC 2075 -- 390740016
and support by the Stuttgart Center for Simulation Science (SimTech).

\vspace{\baselineskip}
\noindent\textbf{Author Contributions}
Z.Z. conducted the model training and analysis. Z.Z.,
H.W., and H.X. contributed to the methodology and manuscript
preparation. Y.K. generated the LES data for the \(7\text{-}7\text{-}7\)
jet-in-crossflow case. H.X. and S.J. supervised the research.
All authors reviewed and approved the manuscript.

\vspace{\baselineskip}
\noindent\textbf{Data and Code Availability}
The solvers, weight-construction scripts, and selected case set-ups supporting
this study are available at:
\begin{center}
\url{https://github.com/ITLR-DDSim/goal-oriented-stress-weighting}
\end{center}
Other supporting data are available from the corresponding authors upon
reasonable request.

\vspace{\baselineskip}
\noindent\textbf{Competing Interests}
The authors declare no competing interests.

\end{document}